\documentclass[alpha-refs]{wiley-article}

\usepackage{graphicx}
\usepackage[labelfont={small}, font={normal}]{subfig} 
\usepackage[space]{grffile}
\usepackage{latexsym}
\usepackage{textcomp}
\usepackage{longtable}
\usepackage{tabulary}
\usepackage{booktabs,array,multirow}
\usepackage{amsfonts,amsmath,amssymb}
\usepackage{natbib}
\usepackage{url}
\usepackage{hyperref}
\hypersetup{colorlinks=false,pdfborder={0 0 0}}
\usepackage{etoolbox}
\makeatletter
\patchcmd\@combinedblfloats{\box\@outputbox}{\unvbox\@outputbox}{}{%
  \errmessage{\noexpand\@combinedblfloats could not be patched}%
}
\makeatother
\newif\iflatexml\latexmlfalse

\AtBeginDocument{\DeclareGraphicsExtensions{.pdf,.PDF,.eps,.EPS,.png,.PNG,.tif,.TIF,.jpg,.JPG,.jpeg,.JPEG}}

\usepackage[utf8]{inputenc}
\usepackage[english]{babel}

\usepackage{siunitx}

\iflatexml
\else

\graphicspath{{./figures/}}

\corraddress{Arjun Sharma, Sibley School of Mechanical and Aerospace Engineering, Cornell University, Ithaca, NY, 14853, USA}
\corremail{as3833@cornell.edu}
\fundinginfo{Airbus Group Corporate Foundation Chair in Mathematics of Complex Systems}
\fi

\papertype{Research Article}

\title{Spatio-temporal relationships between rainfall and convective clouds during Indian Monsoon through a discrete lens}

\author[1]{Arjun Sharma}
\affil[1]{Sibley School of Mechanical and Aerospace Engineering, Cornell University, Ithaca, NY, 14853, USA}		
\author[2]{Adway Mitra}
\affil[2]{Centre of Excellence in Artificial Intelligence, Indian Institute of Technology Kharagpur, India}
\author[3]{Vishal Vasan}
\affil[3]{International Centre for Theoretical Sciences (ICTS-TIFR) Shivakote, Hesaraghatta Hobli, Bengaluru 560089, India}
\author[3]{Rama Govindarajan}

\begin{document}
\maketitle
\selectlanguage{english}
\begin{abstract}
The Indian monsoon, a multi-variable process causing heavy rains during June-September every year, is very heterogeneous in space and time. We study the relationship between rainfall and Outgoing Longwave Radiation (OLR) -– a proxy for convective cloud cover -– for monsoon between 2004-2010. To identify, classify and visualize spatial patterns of rainfall and OLR we use a discrete and spatio-temporally coherent representation of the data, created using a statistical model based on Markov Random Field. Our approach clusters the days with similar spatial distributions of rainfall and OLR into a small number of spatial patterns. We find that eight daily spatial patterns each in rainfall and OLR, and seven joint patterns of rainfall and OLR, describe over 90\% of all days. Through these patterns, we find that OLR generally has a strong negative correlation with precipitation, but with significant spatial variations. In particular, peninsular India (except for the west coast) is under significant convective cloud cover over a majority of days but remains rainless. We also find that much of the monsoon rainfall co-occurs with low OLR, but some amount of rainfall in Eastern and North-western India in June occurs on OLR days, presumably from shallow clouds. To study day-to-day variations of both quantities, we identify spatial patterns in the temporal gradients computed from the observations. We find that changes in convective cloud activity across India most commonly occur due to the establishment of a north-south OLR gradient which persists for 1-2 days and shifts the convective cloud cover from light to deep or vice versa. Such changes are also accompanied by changes in the spatial distribution of precipitation. The present work thus provides a highly reduced description of the complex spatial patterns and their day-to-day variations, and could form a useful tool for future simplified descriptions of this process.

\textbf{Keywords} --- Observational data analysis, Statistical methods, Mesoscale, Seasonal, Clouds, Radiation, Rainfall, Atmosphere%
\end{abstract}

\section{Introduction}\label{sec:Introduction}
The South Asian monsoon season is active from June to September, bringing heavy rains to India, Pakistan, Sri Lanka, Bangladesh, Burma, Nepal and other neighbouring countries. According to present wisdom, during summer, precipitation over a zonal spread including the Indian landmass is caused by the northward shift of a giant mass of clouds called the Inter-Tropical Convergence Zone (ITCZ)~\cite{schneider2014migrations,gadgil2003, sikka1980}. Outside of the Indian Summer monsoon season and the Indian Ocean region, it is wrapped around the Equator. In this paper, we focus on the dynamics of the monsoon over the landmass of India and the surrounding seas (the Bay of Bengal and the Arabian Sea). On one hand, devastating floods result due to excess rainfall as it forces the sudden release of water from a large number of dams and reservoirs, and on the other hand only 10\% less rain than average leads to severe droughts~\cite{gadgil2006}. The Indian Summer monsoon rainfall (ISMR) has a profound impact on agriculture, and hence the economy of India~\cite{gadgil2006}. Hence, understanding the characteristics of the process is of utmost importance. Several aspects of this process, such as strong coupling between ocean and atmosphere processes \cite{wang2005fundamental, pottapinjara2016relation} and absence of any long term trends in ISMR over several decades \cite{goswami2005enso}, make it a challenging problem. 

It is known that a typical season of the Indian monsoon has intra-seasonal oscillations, which are manifested as ``active spells" of high rainfall activity across the landmass, and ``break spells" where most of the landmass remains dry~\cite{rajeevan2006}.  Such intra-seasonal oscillations causes day-to-day changes in the spatial distribution of cloud cover and rainfall. Spatial patterns of daily rainfall across the landmass have been studied using Empirical Orthogonal Functions and their variants~\cite{krishnamurthy2007,mishra2012, suhasMISO}. \citet{krishnamurthy2007} used multichannel singular spectrum analysis of the daily rainfall anomalies and identified a seasonal component and two major sub-seasonal components corresponding to active and break spells. ~\citet{Rajeevanspells} empirically defined active and break spells and spatial distribution of rainfall around them. ~\citet{suhasMISO} identified eight phases of the monsoon, each characterized by a spatial pattern of rainfall, appearing in near-cyclic order in each monsoon season. 

The northward propagation of ITCZ from the equator to the tropical region as the cause of monsoon is well established. The seminal paper by \citet{sikka1980} theorized that there are waves of northward propagation of cloud bands across the Indian landmass. They demonstrated it using Hovm\"oller diagrams along specific longitudes. ~\citet{chattopadhyay2009} also identified northward propagation of the stratiform component of rainfall (not the convective component), through Hovm\"oller diagrams around active and break spells. Several studies have considered the role of clouds, especially synoptic-scale bands during monsoon. Though it is difficult to measure or quantify clouds, Outgoing Longwave Radiation (OLR) is a useful proxy for convection, which plays a major role in the Indian monsoon. It is readily available from many satellites. The propagation patterns of OLR anomalies were studied by \citet{krishnamurthy2008}, who once again tied their analysis around active and break spells to visually demonstrate northward propagation. They also performed spectral analysis of the OLR time-series to identify two pairs of north-south oscillatory patterns of OLR anomalies. They too used Hovm\"oller diagrams to identify a small eastward component and a considerable northward component of the propagation of OLR anomalies over and around the Indian landmass. \citet{rajeevan2013} studied vertical cloud structures during the active and break spells.

We avoid the usual emphasis on ``active'' and ``break'' spells, and instead look to identify typical daily spatial patterns of rainfall and OLR over the Indian region during monsoon. Recently, \citet{mitra2018discrete} identified 10 spatial patterns of rainfall distribution over the Indian landmass during the monsoon season (June-September) on a low-resolution ($1^\circ$) precipitation dataset~\cite{rajeevan2006}. Each pattern is representative of a cluster of days of ``similar'' rainfall distribution, such that each day's spatial distribution of rainfall may be considered to be a noisy version of the pattern of its cluster. Their approach is based on Markov Random Field (MRF) that creates a discrete representation of any spatio\textendash temporal dataset while maintaining spatio-temporal coherence among the discrete state values. Each discrete state represents a probability distribution over the space of observations. The model then partitions the dataset into spatial and temporal clusters that maintain spatio\textendash temporal coherence. \citet{mitra2018spatio} identified temporal relationships between the patterns corresponding to each cluster. The number of clusters is estimated implicitly by the model and studying the corresponding patterns allows new physical insights to be inferred. In this paper, we use the same model as a tool to obtain clusters of patterns on high-resolution satellite-based precipitation (TRMM) and Outgoing Longwave Radiation data over and around the Indian landmass to obtain the spatial patterns of convective cloud and rainfall over the region. We also extend the \citet{mitra2018discrete}  model to detect joint patterns of rainfall and OLR, to explain the relationship between convective cloud cover and rainfall. To the best of our knowledge, this is the first attempt at identifying joint spatial patterns of OLR and rainfall. Our approach identifies eight prominent spatial patterns of precipitation, eight prominent spatial patterns of OLR and seven prominent joint spatial patterns of both OLR and precipitation. We then study their inter-relationship. While low OLR is frequently associated, as would be expected, with high rainfall, and vice versa in most places, we find that this does not hold in the south-eastern peninsula, where low OLR (heavy cloud cover) does not cause significant rainfall.

Our approach to identify spatial OLR patterns indicates that patterns associated with increased convective cloud activity over the entire spatial region studied are positively correlated with increased precipitation over the landmass. To investigate the temporal variation of convective clouds and rainfall we computed one-day anomalies for OLR and precipitation and once again employed our framework to determine spatial patterns of the one-day anomaly. Considering transition frequencies of OLR spatial patterns conditioned on different one-day anomaly patterns, we identified three families of convective cloud cover: high, low and transition. The one-day OLR anomaly analysis also indicates the most common means of shifting convective cloud  activity across India is by establishing a north-south one-day OLR anomaly gradient. The direction of the gradient dictates the direction of the shift in convective activity, either high to low or low to high. Due to the association of higher convective cloud activity with higher precipitation, the direction of the one-day OLR anomaly gradient indicates a period of higher/lower precipitation.

\section{Data and Methodology}\label{sec:Method}
In this section, we briefly discuss the model, based on MRF, proposed by \citet{mitra2018discrete}, which is used to generate discrete spatial patterns of rainfall and OLR. We present only the necessary features for this study and for further details we refer the reader to \citet{mitra2018discrete}. 

The model replaces the observation $X(s,t)$ at any spatial location $s$ and day $t$ by an integer $Z(s,t)$. For example, in the joint patterns of rainfall and OLR, $X$ is a $2\times 1$ vector containing daily rainfall and daily OLR data, and $Z$ is $1$ for low OLR  and rainy, $2$ for low OLR  and dry, $3$ for high OLR  and dry and $4$ for high OLR and rainy. Note that we use the terms ``wet" and ``rainy" interchangeably. The model designs $Z$ for spatial and temporal coherence, i.e. neighboring locations are likely to have the same value of $Z$. Thus for each day during the study period, we have a spatial pattern of $Z$ over the geographical domain. The model collects days with ``similar" $Z$ patterns and puts them into one cluster. A cluster is represented by a discrete spatial pattern (denoted by a vector $\theta_d$) which is the mode of the spatial distribution of $Z$ across its constituent days, and a continuous spatial pattern (denoted by a vector $\theta$) which is the mean of the spatial distribution of $X$ across its constituent days. Each cluster is identified by an integer $U$, and hence each day belongs to a given value of $U$ as elaborated below.

\subsection{Methodology}
Let there be $S$ spatial locations (e.g. a rectangular grid system) and $T$ time-points (here these are days). $X(s,t)$ denotes the observation of any continuous field (e.g. rainfall or OLR) at location $s\in[1,S]$ and time $t\in[1,T]$. 

For every observation $X(s,t)$ (which is a vector for the joint patterns but a scalar for the individual patterns), as mentioned above, the model provides a discrete scalar variable $Z(s,t)$. Each value of $Z$ is the index to a probability distribution over the range of $X$. At most locations, the distributions of rainfall and OLR over time are bimodal, with distinctive peaks. The rainfall distribution at each location is modelled with two Gamma distributions and the OLR with two Gaussian distributions corresponding to lower and higher ranges of the respective variable. In this study, the parameters defining these distributions are made location-specific, but the model allows them to be time-specific as well. In the individual representation for OLR or rainfall at each $(s,t)$ value the variable $Z(s,t)$ can take one of the two possible values (high or low), but in the joint OLR-rainfall representation, one of the four possible values for $Z(s,t)$, as explained above, are possible. 

The model has a discrete variable $U(t)$ for each day, indicating the cluster membership of the day. Each $U$ corresponds to a spatial pattern of $Z$, represented by an $S {\times} 1$ discrete vector, $\theta_d$. These vectors are called {\it spatial patterns}, and the $Z$-vector of each day $t$ is equal (except for a small number of locations) to any one of the spatial patterns.

We now estimate the parameters of the distributions of the model, the spatial patterns ($\theta_d$ vectors) and also find appropriate values of all the $Z$ and $U$ variables. In doing so, the model ensures that $Z$-variables are spatio-temporally coherent, i.e., $Z(s,t)$ is likely to be equal to $Z(s',t')$ if $s' \in \Omega(s)$ or $t' \in \{t-1,t,t+1\}$, where $\Omega(s)$ is the set of neighboring locations of $s$.  Also, the number of spatial patterns is not fixed beforehand but the model finds an appropriate number from the data. They are chosen in such a way that {\it prominent spatial patterns}, which appear repeatedly and in multiple years are identified. These are achieved by considering $Z$ and $U$ as random variables with prior distributions on them, and also their joint distributions specified by pairwise {\it edge potential functions} as per the convention of Markov Random Fields. The optimal values of all the random variables and parameters are obtained by a procedure known as Gibbs Sampling, which is based on Markov Chain Monte Carlo Methods.

\subsection{Datasets}
We study the monsoon months (June- September) from 2004 to 2010. The geographical region analysed is between $7^\circ$N- $38^\circ$N and $68^\circ$E- $100^\circ$E with a resolution of $0.25^\circ$ in either direction. However, to collect daily patterns into clusters, only a region roughly coinciding with the Indian landmass is considered. For precipitation, we use data from the Tropical Rainfall Measurement Mission (TRMM). Our data-set consists of 5772 spatial locations and 854 days.  These 5772 spatial locations are restricted to the roughly the Indian and Bangladesh landmass, i.e., the coloured regions in figure \ref{fig:Discrete Rainfall} or any of the other discrete patterns. The rainfall activity over the Bay of Bengal is much more intense than on the landmass. Thus we have ignored the surrounding seas to allow the clustering process to bring out salient features within the landmass as these features are otherwise subdued in the clustering process. Although the surrounding seas have been ignored during the clustering process, we include these locations in the continuous patterns in figures \ref{fig:PPT_Cluster_PPT_continuous}, \ref{fig:OLR_Cluster_OLR_continuous}, \ref {fig:Joint_Cluster_PPT_continuous}, \ref{fig:Joint_Cluster_OLR_continuous} and \ref{fig:OLRgradcont}. From these figures the spatial features, i.e., regions of high/ low rainfall or OLR appearing over landmass can be observed to smoothly extend over the surrounding oceans. On repeating the experiments with daily precipitation data compiled by India Meteorological Department (IMD, \citet{pai2014}) we find similar results (not shown) as those based on TRMM.

One of the best-known proxy variables for convective clouds is OLR, and such clouds play an important role in the Indian monsoon. Several works such as ~\citet{sikka1980,krishnamurthy2008,utsav2017,chakravarty2018} have used this proxy in the context of the Indian monsoon. Moreover, it is readily available from many earth monitoring satellites. We use the OLR data measured on-board the Kalpana satellite \cite{mahakur2013high} with a resolution of $0.25^\circ$ in either direction. The OLR data-set consists of the same 5772 spatial locations as the precipitation data but only 821 days (data on some of the days during 2004-2010 monsoon months is not available). We consider these 821 days for both the rainfall and OLR clustering.

\section{Spatial Patterns of OLR and Rainfall}\label{sec:ThePatterns}

By applying the Markov Random Field model described in section \ref{sec:Method} on rainfall and OLR datasets,  we obtain three types of discrete spatial patterns: precipitation (PPT) patterns, OLR patterns and joint PPT-OLR patterns. The first two are obtained by using rainfall and OLR respectively as the observed variables for the model, and the discrete latent variable $Z$ takes on one of two values, for ``wet'' and ``dry'' in rainfall or ``deep clouds'' and ``shallow/no clouds'' in OLR. But in the third case, the observed variable is a 2-dimensional vector with both rainfall and OLR. Similarly, the discrete latent variable has 4 states rather than 2.

\subsection{Rainfall and OLR Patterns (obtained separately)}\label{sec:individualpatterns}
\begin{figure}[h!]
	\centering
	\includegraphics[width=5.5in, height=3in]{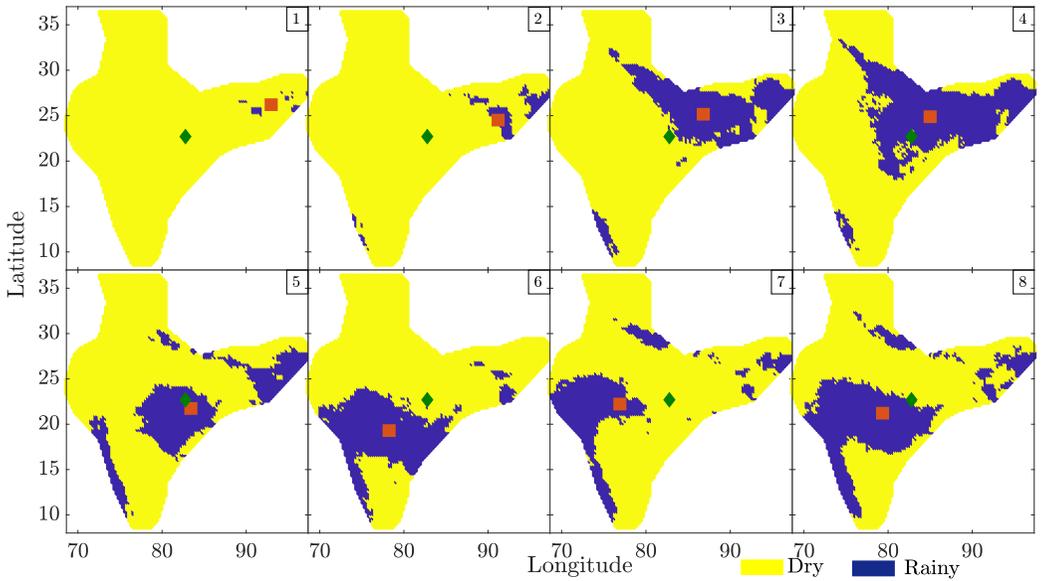}
	\caption {Discrete rainfall patterns arranged in increasing order of mean daily rainfall. Orange square markers indicate the Centre of Mass (Def. 2 in the supplementary material) specific to the pattern and the green diamonds are the Centre of Mass (Def. 2  in the supplementary material)  across all days. The green diamond is hence a fixed reference point against which we compare the migration of the centre of mass for each pattern of rain. \label{fig:Discrete Rainfall}}
\end{figure}

We obtain eight ``prominent" spatial rainfall patterns by applying the model to rainfall data. By ``prominent", we imply patterns which appear at least over 20 days across the seasons on which the model is trained. These patterns, denoted by R1 to R8, are shown in figure \ref{fig:Discrete Rainfall}, where the locations marked in blue are ``wet", i.e., satisfying $\theta_d(s)=1$, while those marked in yellow are ``dry", satisfying $\theta_d(s)=2$. These patterns are similar to the ones obtained by \citet{mitra2018discrete}. The small differences are due to the difference in the data source. As mentioned in Section \ref{sec:Method}, we use satellite-based TRMM data for precipitation, while \citet{mitra2018discrete} had used ground-sensed data published by Indian Meteorological Department. TRMM data is not only of higher spatial resolution but also includes Bangladesh here unlike the other dataset which is restricted to the political borders of India.

\begin{figure}[h!]
	\centering
	\includegraphics[width=5.5in, height=3in]{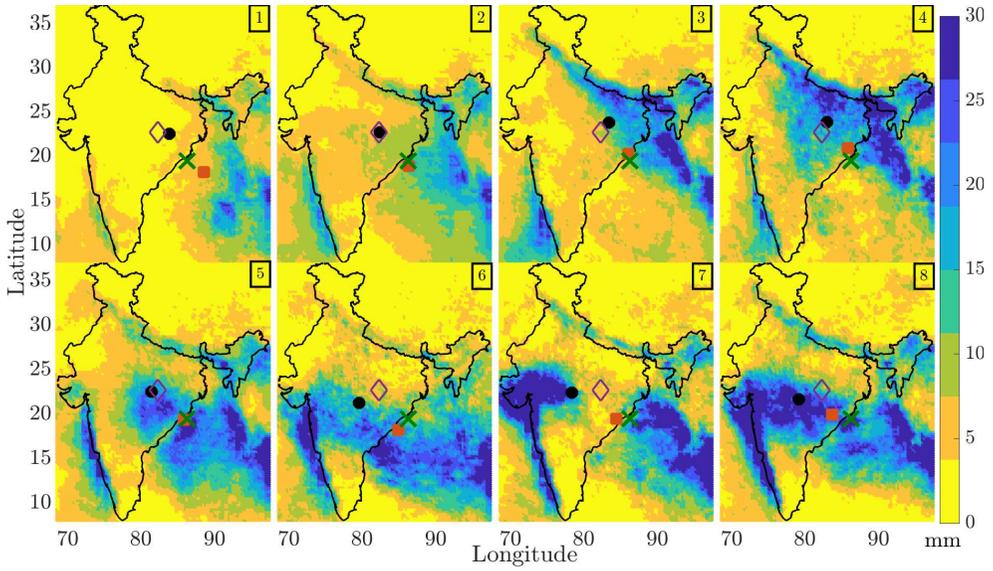}
	\caption {Continuous Rainfall patterns corresponding to the discrete rainfall patterns of figure \ref{fig:Discrete Rainfall}. Black circles and red squares are the Centre of Mass (Def. 1 in the supplementary material) of each pattern's rainfall over the Indian landmass and the full area respectively. The purple diamond and green crosses are the Centre of Mass (Def. 1 in the supplementary material)  of rainfall over the Indian landmass and the full area respectively, across all the days. \label{fig:PPT_Cluster_PPT_continuous}}
\end{figure}

\begin{table}[h!]                           
	\centering
	\begin{tabular}{|c|c|c|c|c|c|c|c|c|}
		\hline
		Rainfall pattern number    &   R1 & R2& R3 & R4& R5& R6& R7 & R8               \\ \hline
		Number of days &  160 & 279& 90 &42& 96& 35& 37 & 42           \\ \hline    
		Mean landmass rainfall  (mm/day) &  4.39 & 6.42& 7.47 &8.31& 8.67& 8.83& 9.07 &10.04 \\ \hline    
	\end{tabular}
	\caption{Number of attributed days over seven monsoon seasons and the daily mean rainfall corresponding to the discrete rainfall patterns of figure \ref{fig:Discrete Rainfall}. These patterns account in total for 781 of the 854 monsoon days in the seven seasons, which is over $91\%$. The rest of the days are classified into non-prominent patterns.} \label{tab:Discrete Rainfall}         
\end{table}
Among the eight patterns in figure \ref{fig:Discrete Rainfall}, the first two only have a few wet locations in the north-east, while in the remaining six there are a considerable number of wet locations across the landmass. All of these six patterns have several wet locations along the coast in the south-west. Patterns R3 and R4 have most of their rainy locations in the north-eastern part of the landmass, including the Gangetic plain. Pattern R5, where parts of the west coast, east-central India (around Odisha) and north-east are wet at the same time, was not observed by \cite{mitra2018discrete}. We attribute this to the fact that we employ a different data set from that used earlier. Pattern R6 to R8 involve rain mostly on the west coast and central India with small parts of north-east and northern India being wet. Table \ref{tab:Discrete Rainfall} shows the number of days in the period 2004-2010 (June-September) which were assigned to these eight patterns, as well as the mean aggregate rainfall (daily, across the landmass) associated with these patterns. More than 91$\%$ of all the days in these seven seasons are classified into one of the patterns in figure \ref{fig:Discrete Rainfall}. The remaining 9$\%$ of days display rainfall patterns that are not prominent, i.e. they appear in only a minority of the seasons. It is clear that the maximum number of days are in the first two ``dry" patterns, while the ``Gangetic Plain patterns" R3 and R4 are slightly more frequent than the ``Central India patterns" R6, R7 and R8. These discrete rainfall patterns correlate well with the corresponding continuous rainfall patterns shown in figure \ref{fig:PPT_Cluster_PPT_continuous}. These continuous patterns show the mean rainfall quantity at each location, over land or sea, of the days corresponding to a given cluster (or spatial pattern). It can be seen that though our clusters were made by only considering landmass rain, each cluster (or pattern) corresponds to a characteristic pattern over the Bay of Bengal. Broadly, higher landmass rain corresponds to higher activity over the Bay. Remarkably the patterns extend seamlessly into the Bay and the Arabian Sea, indicating that the patterns we have detected are parts of larger weather systems characteristic to each of our patterns of rainfall over land. The composite figures of the discrete rainfall patterns, their corresponding continuous rainfall patterns, and the mean spatial distributions of OLR are shown in figure 1 in the supplementary material where it may be visually observed that regions of low OLR often correspond to regions of high rainfall in each pattern.

\begin{figure}[h!]
	\centering
	\includegraphics[width=5.5in, height=3in]{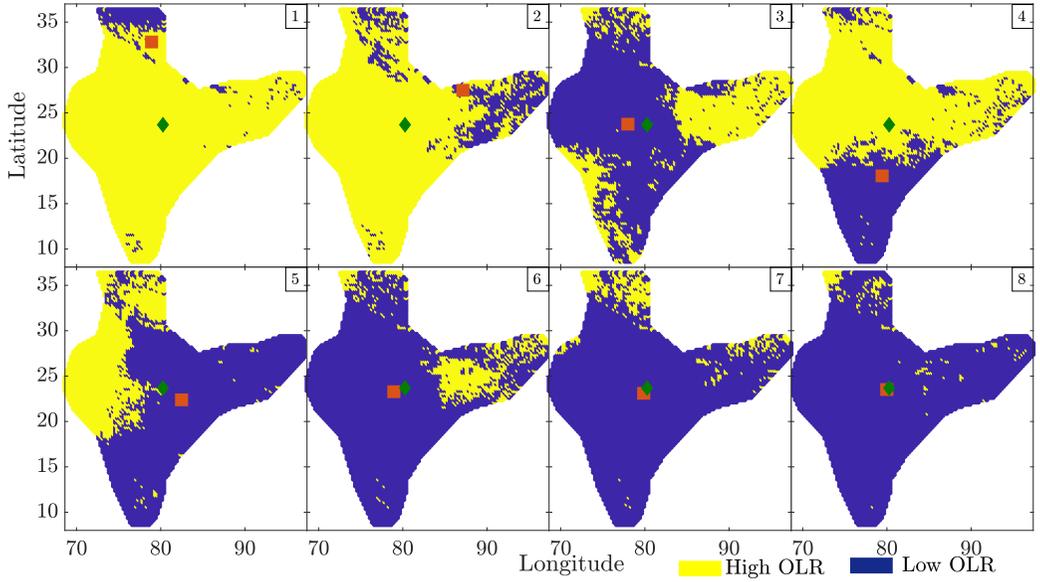}
	\caption {Discrete cloud (OLR) patterns arranged in increasing order of mean daily rainfall. Orange square markers are the Centre of Mass (Def. 2 in the supplementary material)  of high cloud cover locations specific to the patterns and the green diamonds are the Centre of Mass (Def. 2 in the supplementary material) of high cloud cover locations across all days. \label{fig:Discrete OLR}}
\end{figure}

\begin{figure}[h!]
	\centering
	\includegraphics[width=5.5in, height=3in]{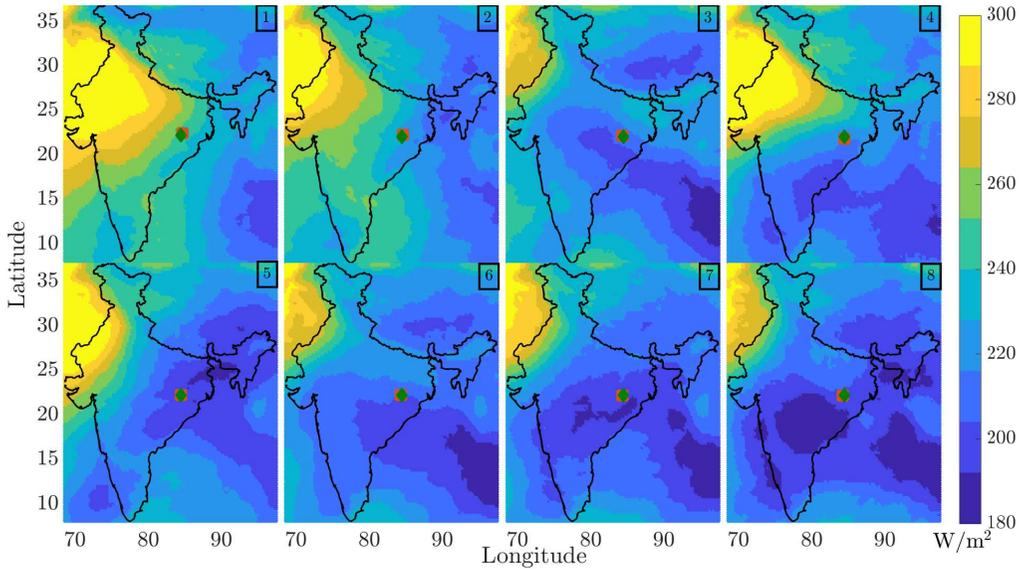}
	\caption {Continuous OLR patterns corresponding to the discrete OLR patterns of figure \ref{fig:Discrete OLR}. Red squares are Centre of Mass (Def. 1 in the supplementary material)  of the full area-weighted with 1/OLR of the pattern. The green diamonds are the Centre of Mass (Def. 1 in the supplementary material) of the full area across all the days sampled.\label{fig:OLR_Cluster_OLR_continuous}}
\end{figure}

Next, we investigate the spatial patterns obtained on OLR data. Once again we obtain eight prominent patterns, O1 to O8, shown  in discrete and continuous forms in figures \ref{fig:Discrete OLR} and \ref{fig:OLR_Cluster_OLR_continuous} respectively. The discrete and continuous patterns correlate well with each other. Since OLR is a proxy for deep convection, the regions with low OLR (shown in blue) are taken to be under dense convective cloud cover, while the regions with high OLR (shown in yellow) have a clear sky or shallow clouds. The patterns are arranged in increasing order of mean rainfall on the days assigned to each OLR cluster. Table \ref{tab:Discrete OLR} shows the number of days during 2004-2010 assigned to each OLR pattern, and the mean aggregate rainfall associated with them. Again the OLR patterns of a vast majority of days in the seven monsoon seasons is represented by these eight prominent patterns, leaving out a negligible number of days which display rare patterns. The first two patterns are mostly devoid of convective clouds and correspond to relatively low rainfall. The other six OLR patterns depict strong convective cloud cover over a significant part of the landmass. Note that low OLR in the northernmost part of the landmass (especially the province of Jammu and Kashmir) may not necessarily correspond to convective cloud cover as this region contains snow-capped mountains. Pattern O4 represents days when primarily the southern part of the landmass is covered by convective clouds. Patterns O3 to O8 show significant presence of convective cloud cover in the south-east, where rainfall during the monsoon season is low. We will return to this point. Patterns O6, O7 and O8 appear similar in their discrete representations, but are progressively more intense in convective cloud cover, as observed in figure \ref{fig:OLR_Cluster_OLR_continuous}.  The composite figures of the discrete OLR patterns, their corresponding continuous OLR patterns, and the mean spatial distributions of rainfall are shown in figure 2 in the supplementary material where once again, it may be visually observed that regions of low OLR often correspond to regions of high rainfall for each pattern.
\begin{table}[h]                           
	\centering
	\begin{tabular}{|c|c|c|c|c|c|c|c|c|}
		\hline
		OLR pattern number    &   O1 & O2& O3 &O4& O5& O6& O7 & O8               \\ \hline
		Number of days &  159 & 107& 67 &86& 102& 159& 103 & 34           \\ \hline    
		Mean landmass rainfall  (mm/day) &  4.40 & 5.70& 6.62 &7.17& 7.28& 8.44& 9.10 &10.44 \\ \hline    
	\end{tabular}
	\caption{Number of attributed days and daily mean rainfall for the discrete OLR patterns of figure \ref{fig:Discrete OLR}. These OLR patterns together account for 817 days (95\% of the monsoon days).} \label{tab:Discrete OLR}                            
\end{table}	

\begin{figure}[h!]
	\centering
	\includegraphics[width=5.5in, height=3in]{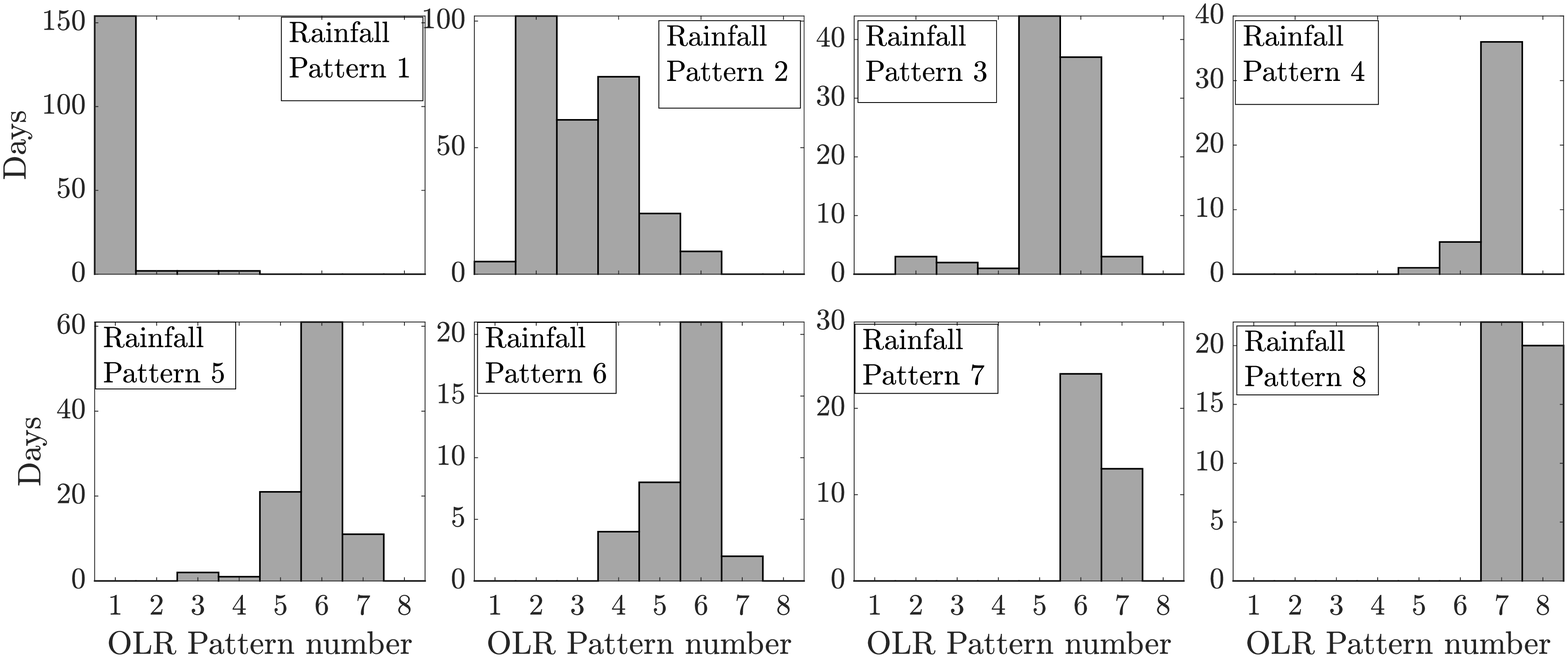}
	\caption {The number of days in which each OLR pattern (figure \ref{fig:Discrete OLR}) occurs for a given rainfall pattern (figure \ref{fig:Discrete Rainfall}). Note the variable vertical axis.} \label{fig:Rainfall_OLR_hist}
\end{figure}

Next, we examine the relationship between the aforementioned rainfall and OLR patterns. Since both sets of patterns have been identified from the same period, each of the 821 days that is assigned to an O-pattern is also represented by one R-pattern. 33 days from R-patterns are left out as OLR data is not available on these days. This allows us to study the correspondence between these two sets of patterns. Histograms of the distribution of OLR patterns across the days attributed to a given rainfall pattern are available in figure \ref{fig:Rainfall_OLR_hist}. The high-OLR pattern 1 in figure \ref{fig:Discrete OLR} corresponds entirely to the driest rainfall pattern R1 in figure \ref{fig:Discrete Rainfall} and vice versa. The rainfall pattern R2, which consists of a small number of wet locations in the north-east, coincides mostly with the OLR patterns O2, O3 or O4, i.e., the ones with patches of low OLR in the north-east. For all the higher rainfall patterns (R3 to R8 in figure \ref{fig:Discrete Rainfall}) the corresponding OLR patterns are seen to be O5 to O8 of figure \ref{fig:Discrete OLR}. From the continuous versions of these rain-providing (to the landmass) OLR patterns in figure \ref{fig:OLR_Cluster_OLR_continuous} we can observe an intense convective cloud band (a trough in OLR) in the south-eastern portion of the Bay of Bengal that partially extends to the landmass. OLR patterns O3 and O4 also share this feature. The convective cloud band in O3 is thinner and more diagonal covering only central India in the landmass part. Pattern O4 has a convective cloud band sitting on the southern part of landmass (also extending to the bay). It mainly coincides with the dry rainfall pattern R2 (it does provide rain in Kerala on a very small number of days through rainfall pattern R6).  Large areas of high OLR in the north-west make this a non-raining cloud pattern. Hence, in most parts of India, there is a clear anti-correlation between OLR and rainfall, but this is not true in southern India where low OLR is often not accompanied by rainfall. This will be a particular feature of the joint patterns discussed in Section \ref{sec:Joint}. 

Rainfall patterns R3 and R4 are qualitatively similar but the corresponding OLR patterns are different. OLR patterns O5 and O6 are responsible for the scanty rainfall pattern R3 and the OLR pattern O7 is responsible for the more intense but rarer rainfall pattern R4 (table \ref{tab:Discrete Rainfall}). Rainfall pattern R5 occurs mostly on the days of OLR pattern O6. OLR pattern O6 is also responsible for rainfall pattern R6 and is a major part for rainfall pattern R7. The most intense rainfall pattern R8 is brought about in almost equal proportion by the OLR patterns O7 and O8. OLR pattern O7 is thus partly responsible for the intense versions of particular rainfall patterns, i.e., rainfall over the north-east via pattern R4 and rainfall over the west coast and central India via pattern R8. Unlike other rain-providing OLR patterns, O7 has two intense peaks in convective cloud cover (or troughs in OLR): one in the south-east Bay and other in central India. OLR pattern O8, which is responsible for rainfall pattern R8 on about 50\% of the days, has three intense peaks of convective cloud cover: in the south-east Bay, over central India and in the North East, but also forms a continuous band from the south-east Bay up to the landmass. In general, the low OLR patterns cover a broader part of the landmass than high rainfall patterns, which implies that on an average, it rains over a subset of cloudy regions, as is reasonable to expect.

\subsection{Joint Patterns}\label{sec:Joint}
\begin{figure}[h!]
	\centering
	\includegraphics[width=5.5in, height=3in]{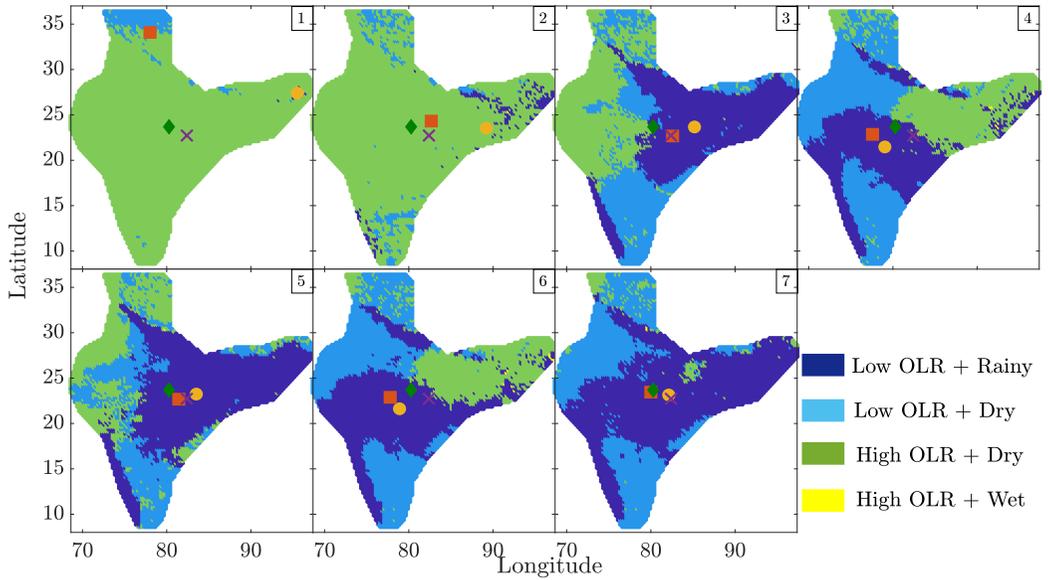}
	\caption {Discrete joint patterns of OLR and rainfall arranged in increasing order of mean daily rainfall. Yellow circle and orange square markers are the centre of Mass (Def. 2 in the supplementary material)  of high rainfall and high cloud cover (low OLR) locations respectively specific to the patterns. The green diamond and purple cross are the Centres of Mass (Def. 2 in the supplementary material) of high rainfall and high cloud cover (low OLR) locations respectively over all days.\label{fig:Joint}}
\end{figure}

Next, we consider the patterns obtained by considering OLR and rainfall together as observations, as discussed in Section~\ref{sec:Method}. Seven prominent patterns, denoted by J1 to J7, where J stands for ``joint", are obtained. The latent variable $Z$ can now take four values (high OLR-high rain, low OLR-high rain, low-OLR-low rain, high OLR-low rain), instead of 2 as in the previous cases. Figure \ref{fig:Joint} shows these patterns, where the four values of Z are suitably color-coded. Table \ref{tab:Joint} shows the number of days in which each joint pattern is displayed.  
\begin{table}[h!]                           
	\centering
	\begin{tabular}{|c|c|c|c|c|c|c|c|}
		\hline
		Joint OLR-rainfall pattern number    &   J1 & J2& J3 &J4& J5& J6& J7                \\ \hline
		Number of days &  156 & 263& 124 &123& 77& 44& 30            \\ \hline    
		Mean landmass rainfall  (mm/day) &  4.39 & 6.38& 7.52 &8.46& 8.71& 9.71& 10.41  \\ \hline    
	\end{tabular}
	\caption{Number of attributed days and the corresponding daily mean rainfall for the discrete joint patterns of figure \ref{fig:Joint}. $817$ of the $854$ monsoon days are classified into these patterns.} \label{tab:Joint}                            
\end{table}

For all the patterns, the number of high OLR-high rainfall locations is insignificant. In patterns J1 and J2, most locations have high OLR and low rainfall. For the patterns with a significant number of rainy locations, i.e., joint patterns J3 to J7 in figure \ref{fig:Joint}, we observe that the west coast is always rainy and cloudy (the wet rainfall patterns of \citet{mitra2018discrete} also had several wet locations on the west coast). Also, southern India is always cloudy in patterns J3 to J7, but includes no rainfall, except for the west coast. Patterns J3 and J5 correspond to convective rainfall in the north-eastern side while the western side remaining uncovered by convective clouds. Similarly, patterns J4 and J6 correspond to a high OLR in north-east, cloudy north-west, and rainy west coast and central India. Pattern J7 corresponds to convective cloud cover all over India with rain over the west coast and the central and north eastern regions. 

\begin{figure}[h!]
	\centering
	\includegraphics[width=5.5in, height=3in]{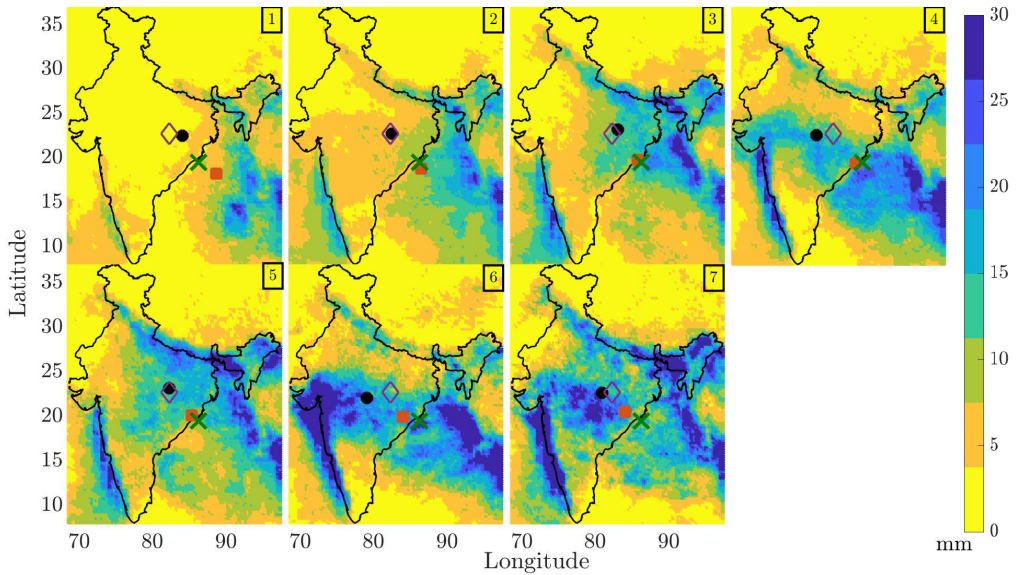}
	\caption {Continuous Rainfall patterns corresponding to the discrete joint patterns of figure \ref{fig:Joint}. Black circle and red squares are Centres of Mass (Def. 1 in the supplementary material)  of the Indian landmass and the full area corresponding to the patterns. The purple diamonds and green crosses are the Centres of Mass (Def. 1 in the supplementary material)  of the Indian landmass and the full area across all the days sampled. \label{fig:Joint_Cluster_PPT_continuous}}
\end{figure}
\begin{figure}[h!]
	\centering
	\includegraphics[width=5.5in, height=3in]{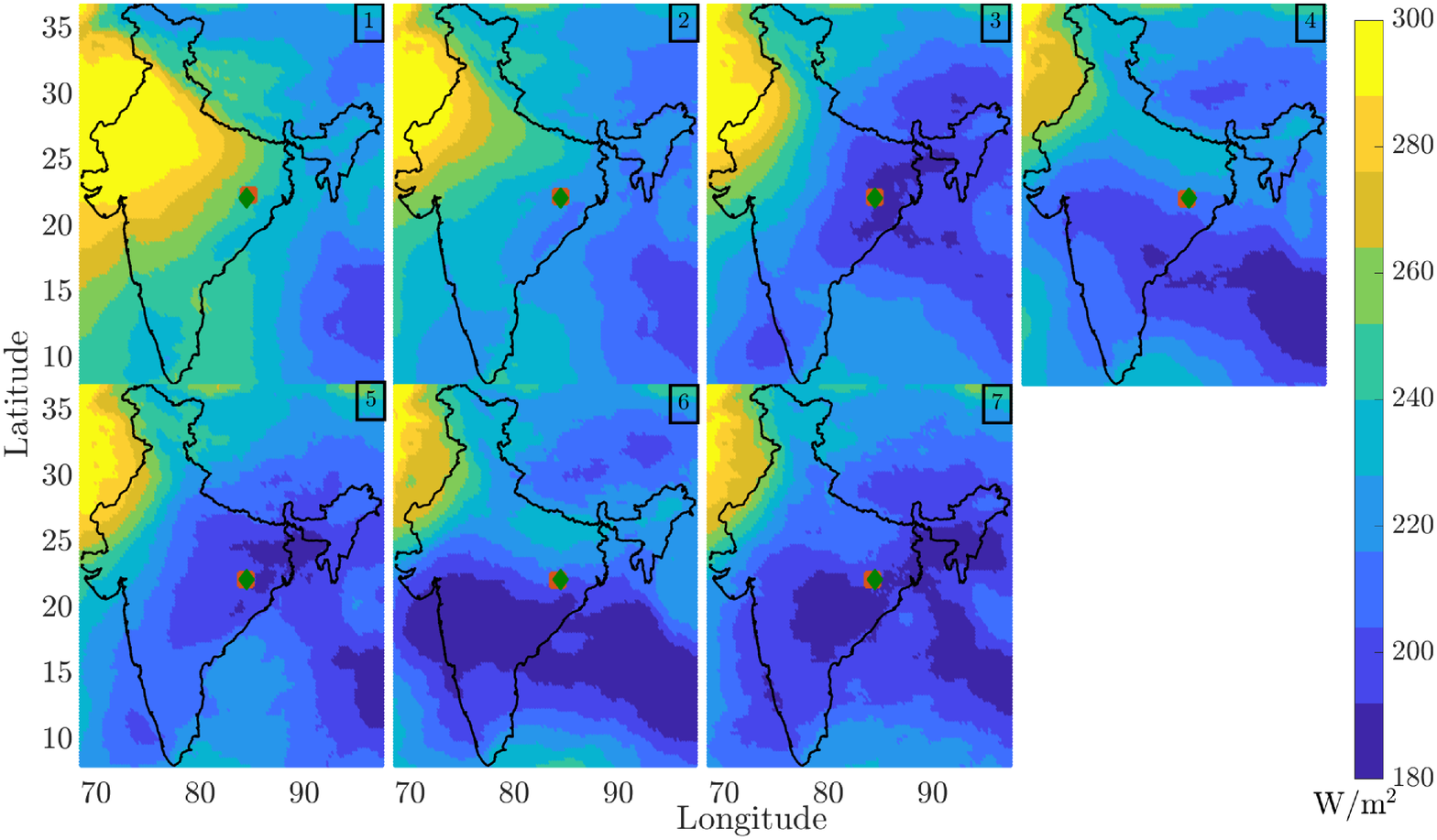}
	\caption {Continuous OLR patterns corresponding to the discrete joint patterns of figure \ref{fig:Joint}. Red squares are Centres of Mass (Def. 2 in the supplementary material)  of the full area-weighted with 1/OLR corresponding to the patterns. The green diamonds are Centres of Mass (Def. 2 in the supplementary material) of the full area across all the days sampled. \label{fig:Joint_Cluster_OLR_continuous}}
\end{figure}

Figures \ref{fig:Joint_Cluster_PPT_continuous} and \ref{fig:Joint_Cluster_OLR_continuous} shown the continuous rainfall and OLR patterns corresponding to the discrete joint patterns of figure \ref{fig:Joint}, where a good correlation between the continuous and discrete information can be observed. The continuous OLR patterns reveal further relation between the specific OLR and rainfall distributions. Joint patterns J3 and J5 which correspond to more rain in the north-east and less in central India, consist of convective cloud bands that are aligned roughly diagonally (from north-west to south-east). The joint patterns J4, J6 and J7 which bring significant rainfall over central India, corresponding to a more horizontal convective cloud band over the same region. The last of these also consists of more intense cloud band over the Bay of Bengal that stretches towards the south-east. There is a separate disjoint convective peak over Kerala along the south-western coast in patterns J3 and J5 in figure \ref{fig:Joint_Cluster_OLR_continuous}. Perhaps this is the topographical effect of the Western Ghats mountain range that stretches along the western coast, causing an increased convective cloud intensity and rainfall in this region even for the diagonally aligned cloud-bands. The composite figures of the discrete joint rainfall and OLR patterns and their corresponding continuous rainfall and OLR patterns are shown in figure 3 in the supplementary material. 

In the literature, it is common to study daily anomalies of rainfall or OLR instead of their absolute values. For consistency with the literature, we also carried out the same analysis using the daily anomaly values of both quantities and obtained very similar patterns as shown above (not shown).

\subsection{Spatio-temporal OLR-rainfall relationship}
At each spatial location, we now identify the probability of occurrence of each of the four states, namely LOW (low OLR with high chance of rainfall), LOD (low OLR but no rainfall), HOD (high OLR and dry), and HOW (high OLR and rainy). We count the number of days with each of the four states as computed by the joint analysis. This number divided by the total number of days for all the four states is used to define a 2$\times$2 probability matrix of the joint states at each spatial location. We can estimate these numbers in two ways - either directly from the $Z$-variables whose binary values are assigned by the inference algorithm, or from the joint patterns computed above. The difference between these two approaches is that the second represents the modes, i.e. most frequent behaviors, while the first approach can account for infrequent behaviors also. 

\begin{figure}[h!]
	\centering
	\subfloat{\includegraphics[width=0.5\textwidth]{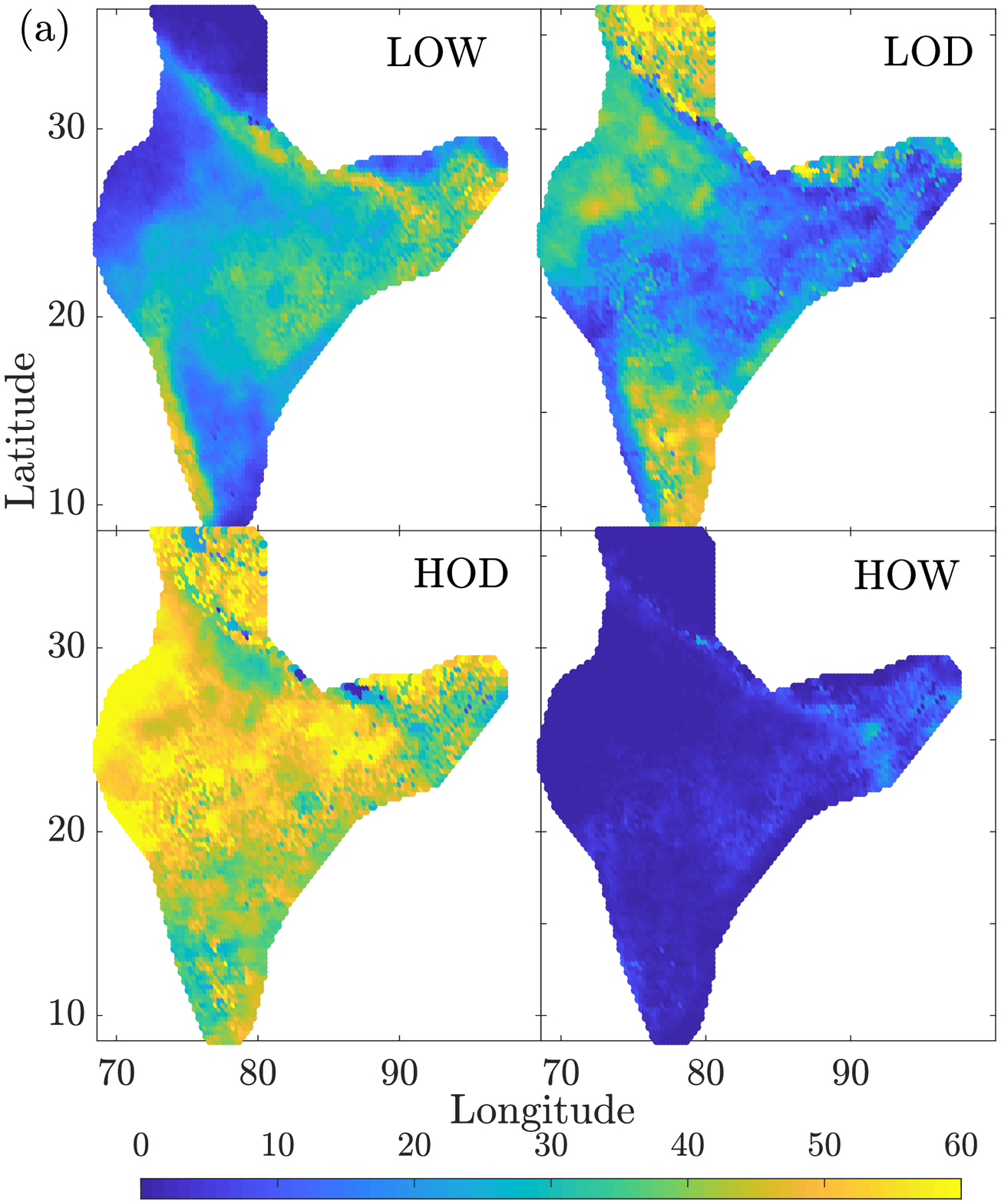}\label{fig:RainyCloudy}}\hfill
	\subfloat{\includegraphics[width=0.49\textwidth]{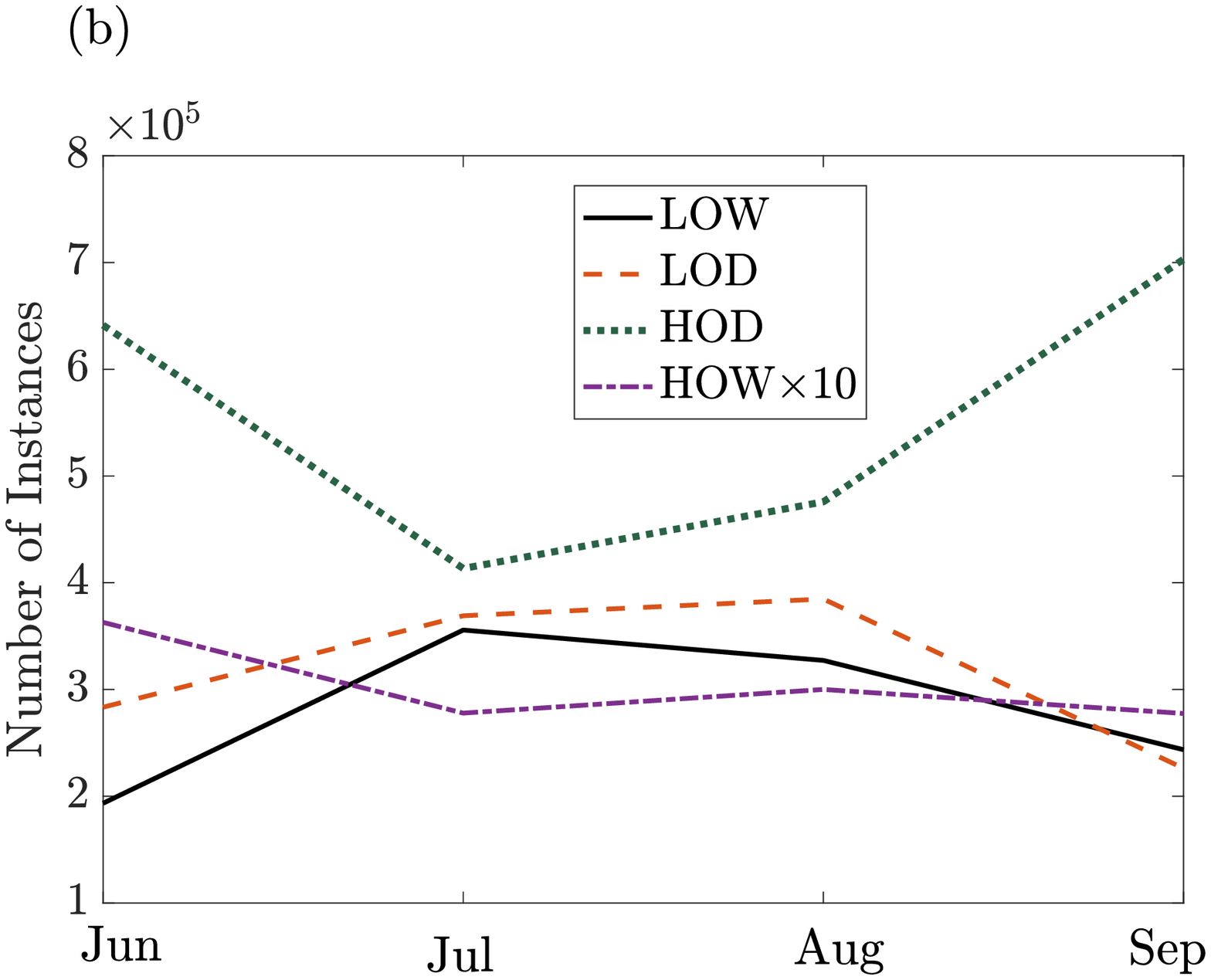}\label{fig:RainyCloudyMonths}}
	\caption  {(a) Percentage of days at each location corresponding to the four OLR-rainfall states: LOW (low OLR and wet), LOD (low OLR and dry), HOW (high OLR and wet) and HOD (high OLR and dry), (b) Total number of monthly instances of each OLR-rainfall state. \label{fig:RAINCLOUD}}
\end{figure}

First, we study the approach of estimating the OLR-rainfall relation directly from $Z$ and $X$. In the period under analysis (7 monsoon seasons spanning 854 days over 5772 locations), there are 23.6\% LOW, 26.7\% LOD, 47.1\% HOD and only 2.6\% HOW events. An event refers to a spatio-temporal point (s,t) which is assigned a states Z(s,t) by the proposed model (with respect to both OLR and rainfall)  before we calculate the mode across the days constituting a cluster (this mode defines the discrete pattern of figure 6). The spatial and temporal distribution of these events are shown in Figure \ref{fig:RAINCLOUD}. 

\begin{figure}[h!]
	\centering
	\subfloat{\includegraphics[width=0.5\textwidth]{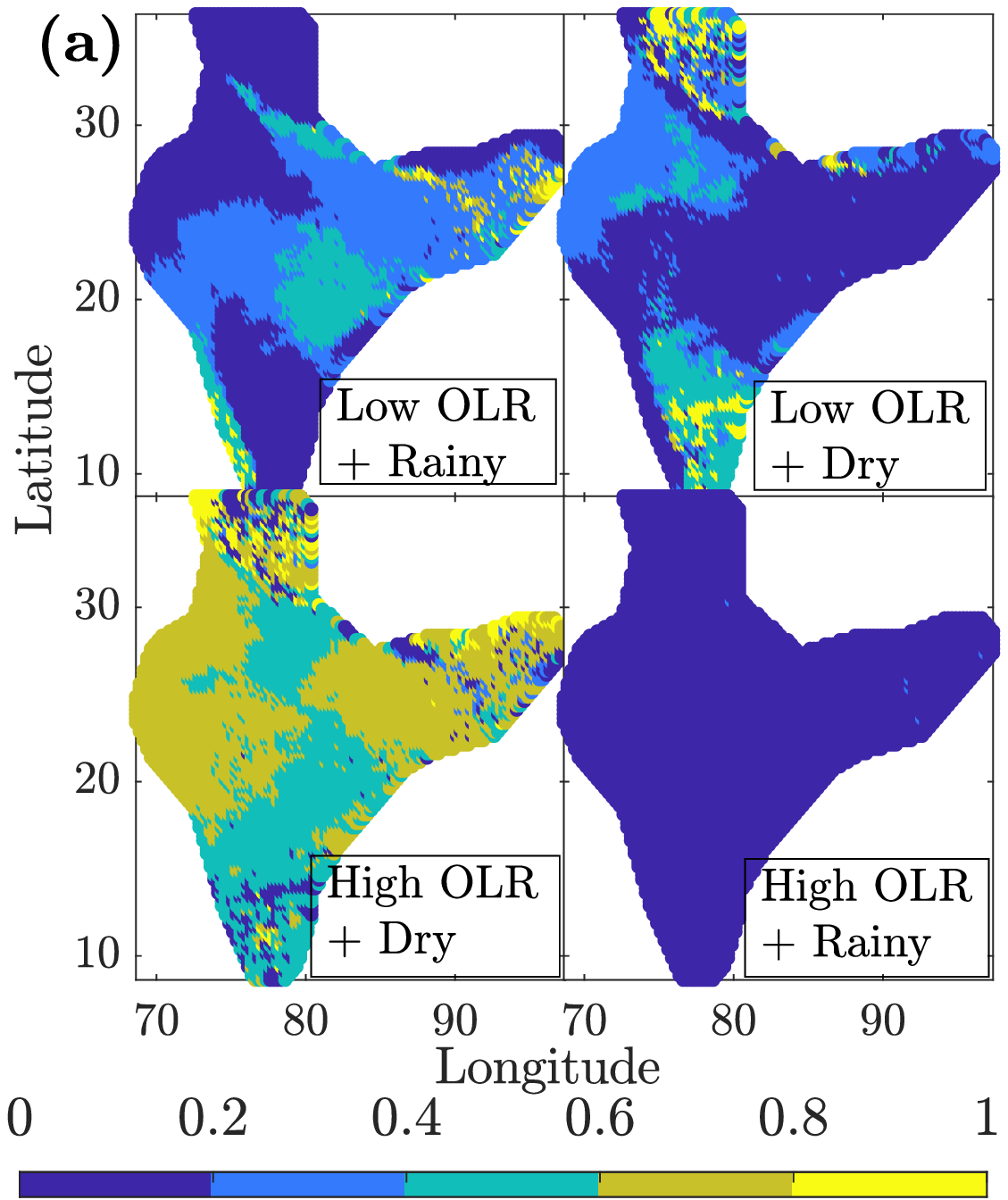}\label{fig:Conditional_joint}}\hfill
		\subfloat{\includegraphics[width=0.4\textwidth]{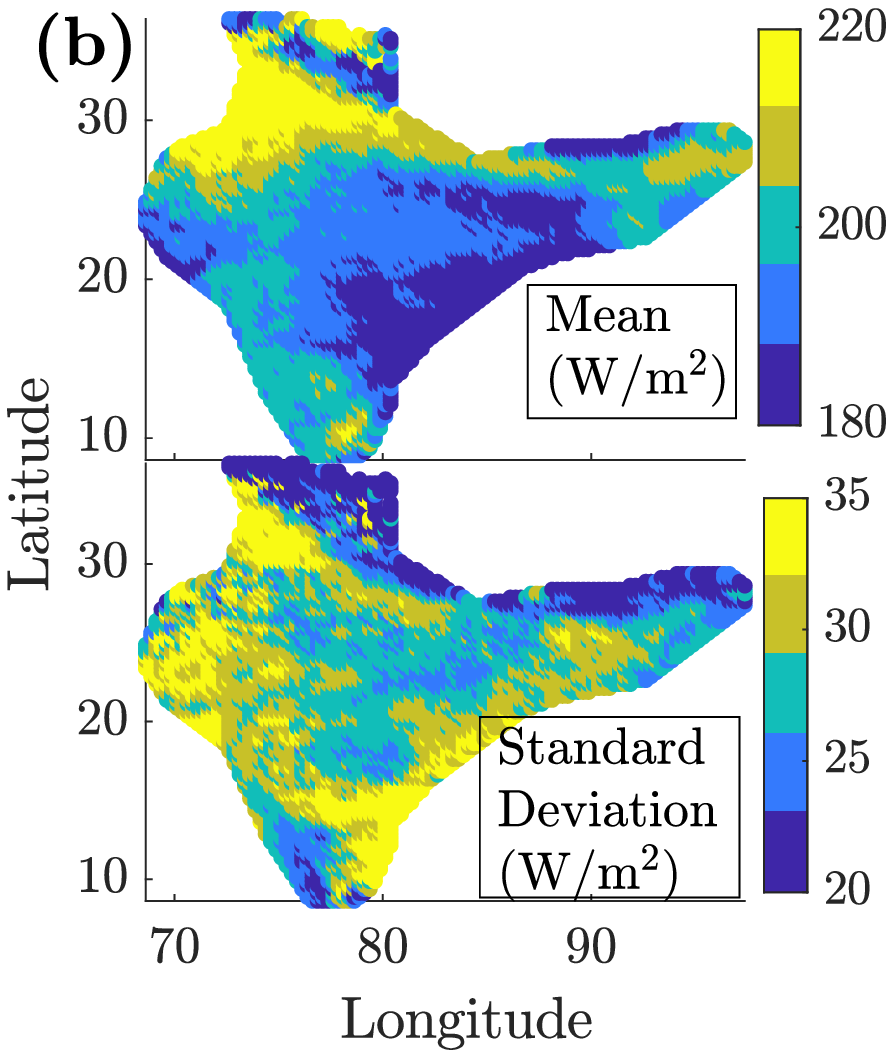}\label{fig:WetOLRStats}}
	\caption {How does OLR relate to rainfall?: a) Conditional matrices from the joint clustering. Yellow corresponds to high likelihood while deep blue indicates a low probability, b) Mean and standard deviation of OLR values on rainy days. \label{fig:Conditional_jointOLRstats}}
\end{figure}

According to the mode-based analysis, each element of the matrix at each location is shown on a map in figure \ref{fig:Conditional_joint}. This provides further credence to observations already made while revealing some new features. Broadly, the landmass may be divided into four regions in terms of this behaviour: Kerala and the Western Ghats; the rest of South India apart from this region; the north east, the Ganga basin, central India; and the northwest. All the regions have a significant but varying probability of being in a high OLR and dry state (``break spells"). In the Kerala and Western Ghats, it is cloudy and rainy with high probability, with a small number of high OLR and dry days in parts. Over the North-East, the Ganga basin and central India, it is almost always either LOW or HOD, with the probability of each varying with location. It is to be remembered from our earlier discussion, however, that the individual days on which a given state occurs are quite different in central India and the North East. The predominant state in the northwest is HOD. Surprisingly, the most likely state in the rest of South India is LOD, with a smaller probability of HOD. Thus, convective clouds most often cover the rest of South India but do not cause rain. The 2.6\% HOW events get suppressed upon taking the mode, and hence do not show up in Figure\ref{fig:Conditional_jointOLRstats}. Although rainfall from non-convective clouds does occur in Indian monsoon, especially in the Western Ghats ~\cite{utsav2017}, this is not a modal feature as it does not appear anywhere in any pattern.

The above analysis makes it clear that rainfall is almost always associated with reduced values of OLR. We know that OLR is a proxy for deep convection which is likely to cause rainfall, but it is possible for local rainfall can occur from shallow non-convective clouds too, which do not cause low OLR. There is no cut-off on OLR values to differentiate between convective and non-convective rainfall.  How may OLR be connected to significant rainfall? The answer may once again vary spatially since rainfall happens due to different mechanisms in different parts of the landmass. At each location, we note raw OLR values for the binary ``wet" days there (as identified by the model). We calculate the mean and standard deviation of these values, and plot them in figure~\ref{fig:WetOLRStats}. It can be observed that generally, rainfall occurs at lower OLR values (180-185 $W/m^2$) along the eastern coast, adjacent to the Bay of Bengal, and parts of the Gujarat coast along the Arabian Sea, while for Central India this value is in the range 190-195 $W/m^2$.  For the long south-western coast and the hilly north-eastern region, this range is even higher, around 200 $W/m^2$, while in North-western India it is even higher, 210 $W/m^2$ or more. However, the variability of these values is quite high along the eastern coast and Gujarat coast, very high in the north-west, but low in the North-eastern region, Central India and the South-western coast. The low mean values of OLR along the eastern coast is consistent with the fact that much of the rainfall there occurs due to deep convective events that happen over the Bay of Bengal, resulting in frequent lows and depressions. The same is not true about much of the western coast, except its northern parts in Gujarat (Kathiawar peninsula).

\subsection{Comparison with known spatial patterns}
\begin{figure}[h!]
\centering
\includegraphics[width=1\textwidth]{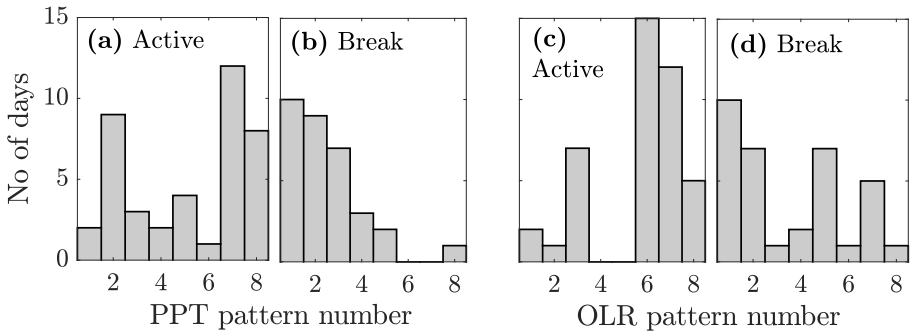}
\caption{Frequency distribution of PPT (a, b) and OLR (c, d) patterns identified by our method during the active (a, c) and break spells (b, d) identified by~\citet{Rajeevanspells}. They supply the list up to 2007. Note that only a minority of days are classified as either active or break.}\label{fig:activebreakhist} 
\end{figure}

Several papers have identified various spatial patterns related to rainfall and other climatic variables related to the Indian monsoon. These include the study of active and break spells by~\citet{Rajeevanspells}, and that of Monsoon Intra-seasonal Oscillation (MISO) by ~\citet{suhasMISO}. In the case of ~\citet{Rajeevanspells}, a list of active and break spells over India is provided, as is the mean spatial distribution of rainfall across the landmass on days leading up to or following such spells. We now ask whether our assignment of each day to a pattern is consistent with the classification of active and break days. We obtain frequency distributions over the patterns for active and break days as identified by~\cite{Rajeevanspells}, and show their histograms in figures~\ref{fig:activebreakhist} for PPT and OLR. We find that most of the ``active" days are assigned to PPT patterns 7 and 8 which correspond to heavy rainfall over central India and the west coast, which is expected during active spells. Similarly, the days identified as ``break" are predominantly assigned to PPT patterns 1 or 2 which are mostly dry. Similar features are visible in the OLR patterns too. Thus our results not only agree with previously known ones but present spatial patterns in a more detailed and quantitative fashion. For example, we have shown that every day's OLR and rainfall are one among a small number of spatial patterns.

\begin{figure}[h!]
\centering
\includegraphics[width=5.5in, height=3in]{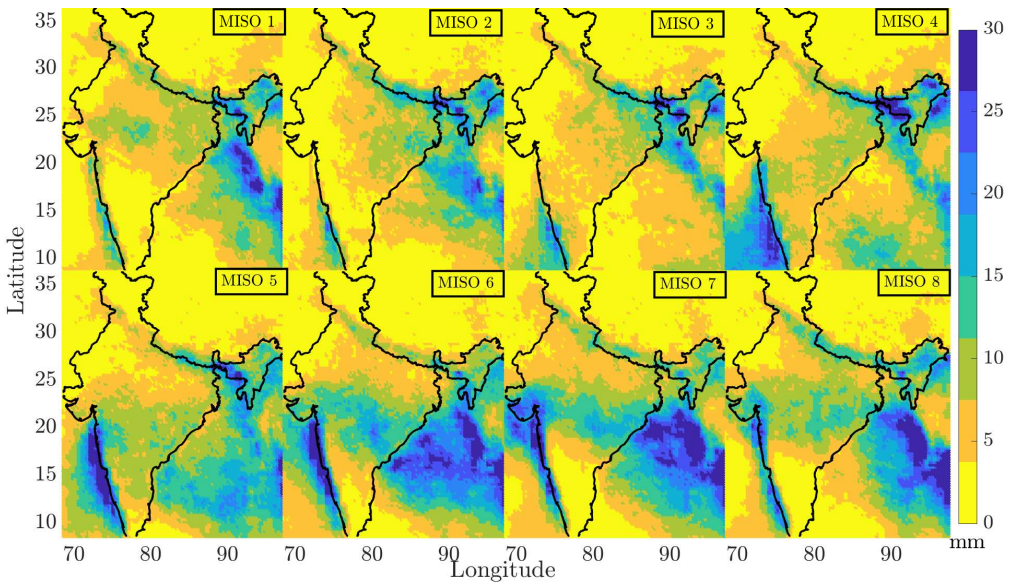}
\caption{Mean rainfall maps of the eight MISO phases identified by ~\citet{suhasMISO}. Each plot is obtained by averaging the rainfall across all days in a given MISO phase.}\label{fig:MISOmaps} 
\end{figure}

The work of ~\citet{suhasMISO} uses Extended Empirical Orthogonal Functions (EEOF) to identify eight phases of monsoon, and each day is assigned to one of those phases, except for those days when the signal is weak. We collect the days in each of their MISO (Monsoon Intra-Seasonal Oscillations) phases, compute the mean spatial rainfall maps of each phase, and show them in figure~\ref{fig:MISOmaps}. While they differ in detail from our rainfall patterns, we see broad similarity in that the eight MISO phases too broadly fall into the three ``families" (similar to those found by \cite{mitra2018spatio}) - i) where the rainfall is restricted to the western coast and the north-eastern regions, ii) rainfall covers central India, iii) rainfall covers eastern India and the Gangetic plains. By counting the pattern assignments from our method in a particular phase of MISO, we find that each pattern roughly overlaps with 3 MISO phases, generally from the same ``family". For example, the 1st pattern overlaps with MISO phases 1,2,3 and in all cases, the rainfall is concentrated in the North-east and foothills of the Himalayas and the western coast. MRF pattern 5 overlaps with MISO phases 6 and 7, where the rainfall occurs mostly in Central India. To put this conclusion from visual comparison on firmer footing, we calculated the correlation coefficient of the spatial rainfall distribution on each day in a particular MISO phase with each of eight PPT patterns ($\theta$) identified by us, and this corroborated the visual comparison.

There are other studies which discuss spatial patterns of rainfall over India, such as \citet{chattopadhyay2008objective} who carried out a clustering based on Self-Organizing Maps to identify the modes of intra-seasonal variations of monsoon, and identified spatial patterns of rainfall anomalies. Some of these spatial patterns have similarity with some of the patterns identified by our model, especially the modes (1,1), (2,1), (3,1) and (3,2) where the rainfall is primarily over Central India or Gangetic plains. However, since their reported results are for an earlier period (1980-1996), we could not make a direct comparison.

\section{Temporal variation in OLR spatial patterns}\label{sec:PropagationMain}

Next, we proceed to investigate the temporal dynamics of convective cloud bands and its consequences on rainfall over the landmass. We first compute daily fluctuations or {\textit{one-day anomalies}} of the OLR and precipitation data as follows: $dX(s,t) = X(s,t) - X(s,t-1)$ , where $X$ denotes any measurement - either precipitation or OLR. A positive value indicates an increase of the variable relative to the previous day, while a negative value indicates a decrease. We then ran our MRF model on this one-day anomaly data, where the binary latent variables $dZ(s,t)$ indicate either positive or negative gradients. The data-edge potential functions are now defined using the sigmoid function as follows:
\begin{equation}
\psi(dZ(s,t),dX(s,t)) = \frac{1}{1+\exp{((-1)^{dZ(s,t)}dX(s,t))}}, dZ\in[1,2].
\end{equation}

\begin{figure}[h!]
	\centering
	\includegraphics[width=5.5in, height=3in]{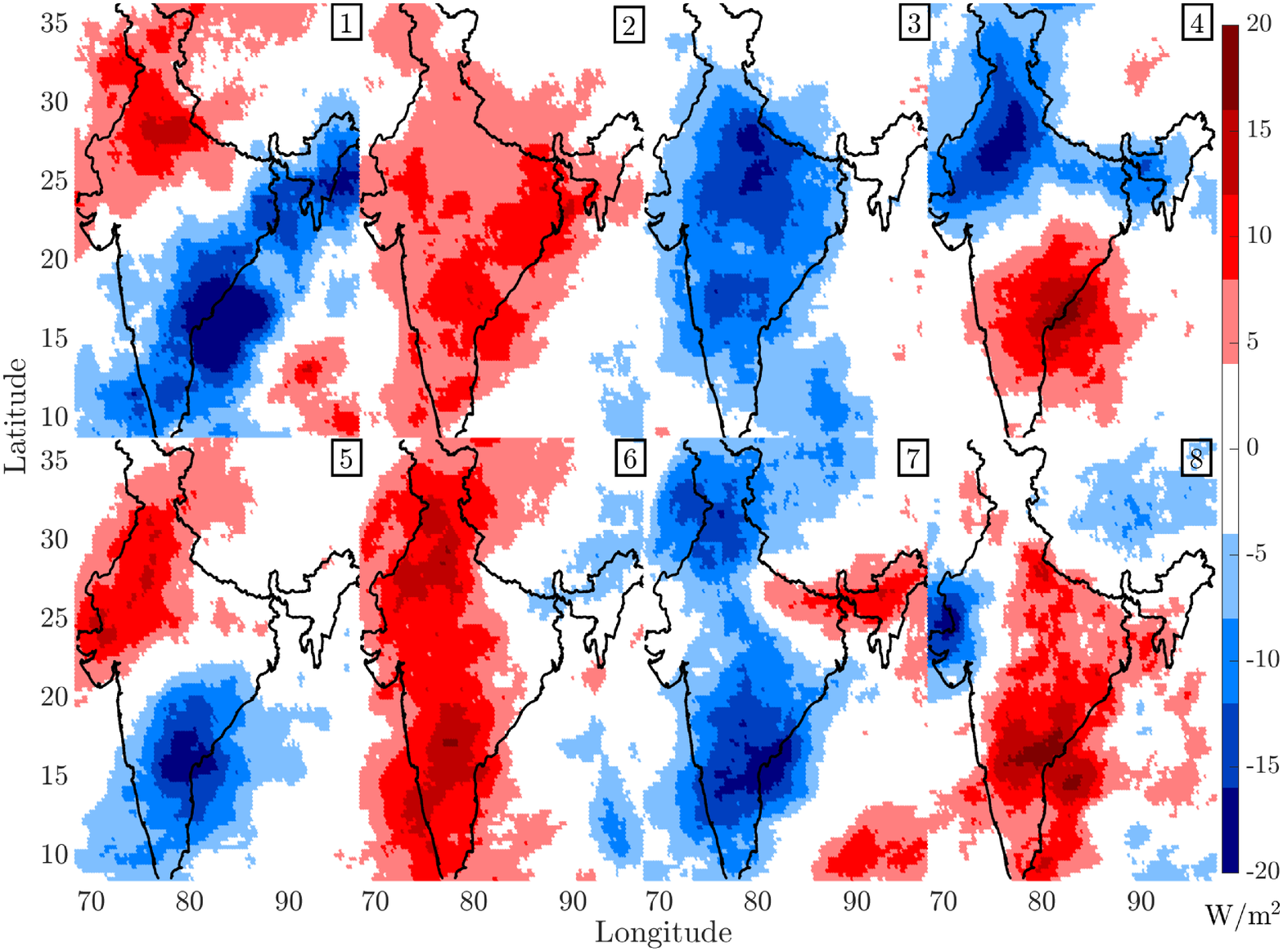}
	\caption {Continuous spatial patterns of one-day OLR anomaly, during the period 2004-10. Red indicates an increase in OLR (less convective cloud on the following day), blue indicates a decrease in OLR (more convective cloud on the following day). \label{fig:OLRgradcont}}
\end{figure}

We find eight prominent patterns in one-day OLR anomaly, each of which appears on at least 35 days during the period under consideration. The continuous spatial patterns of OLR, the mean one-day anomalies for a particular spatial pattern, $\theta$ are shown in figure~\ref{fig:OLRgradcont}. Red indicates that OLR is higher, i.e. less convective cloud cover on the following day than on the current day, and blue indicates vice versa. Their discrete counterparts $\theta_d$ produced by this approach are shown in the supplementary material in figure 4. Frequencies of the patterns and their relationships with daily aggregate rainfall are shown in table~\ref{tab:ContOLRgrad}. Pattern 1 appears on relatively dry days. Patterns 2 and 3 appear mostly in June and September when the convective cloud cover is relatively low and sporadic. Patterns 4 and 5 are complementary, indicating increasing OLR over South India and decreasing over North India, and vice versa. Both patterns 4 and 5 appear in all Monsoon months evenly. Similarly, the patterns 6 and 7 are complementary, one showing increase of OLR all over India except the north-eastern corner, and the other showing the reverse. Pattern 8 shows a decrease of OLR over Western India, with an increase over the rest of the landmass.

\begin{table}[h]
	\centering
	\begin{tabular}{|c|c|c|c|c|c|c|c|c|}
		\hline
		One-day OLR anomaly pattern number & OG1 & OG2& OG3 &OG4& OG5& OG6& OG7 & OG8 \\ \hline
		Number of days per season &5.6 & 15.1 & 12.7& 18.7 &18.1& 16.1& 15.1& 8.4 \\ \hline
		Mean landmass rainfall (mm/day) & 4.4 & 4.7 & 5.2& 6.8 &7.0& 8.0& 9.0& 9.14 \\ \hline
	\end{tabular}
	\caption{Number of attributed days and daily mean rainfall for the discrete rainfall temporal gradient patterns of figure ~\ref{fig:OLRgradcont}. These one-day OLR anomaly patterns together account for 770 of the 854 monsoon days.} \label{tab:ContOLRgrad}
\end{table}

\begin{table}[h]
	\centering
	\begin{tabular}{|c|c|c|c|c|c|c|c|c|}
		\hline
		& O1 & O2 & O3 & O4 & O5 & O6 & O7 & O8 \\
		\hline
		O1 & 2 (45)	& 1 (8) & 3 (1) & 3 (8) & 3 (3) & 3 (1) & 3 (2) & X (0)\\
		O2 & 2 (17)	& 4 (14) & 2 (1) & 5 (9) &	5 (5) & 4 (1) & 3 (4) & X (0)\\
		O3 & 5 (3) & 2 (2) & 4 (10) & 5 (1) &	5 (7) & 4 (13) & 5 (1) & 4 (1)\\
		O4 & 2 (3) & 4 (6) & 4 (5) & 5 (15) & 5 (5) & 4 (3) & 4 (1) & X (0)\\
		O5 & 5 (1) & 6 (7) & 4 (3) & 6 (4) & 5 (15) & 4 (6) & 7 (6) & 4 (1)\\
		O6 & 6 (2)	& 6 (4) & 6 (11) & 6 (4) &	6 (11) & 6 (35) & 7 (16) & 6 (3)\\
		O7 & X (0) & 8 (1) & 8 (4) & 8 (1) & 7 (5) & 8 (19) & 8 (18) & 8 (7)\\
		O8 & X (0)	& X (0) & X (0) & X (0) & X (0) & 8 (2) & 8 (2) & 8 (1)\\
		\hline
	\end{tabular}
	\caption{The most common one-day anomaly pattern of OLR (OG), through which the transition occurs from one OLR spatial pattern (O) to another in successive days. Each row indicates the current day's spatial pattern, each column indicates the following day's spatial pattern, and the entry contains the most frequent one-day anomaly pattern. The frequency of such a transition is shown in brackets. The rows and columns labeled as O1- O8 refer to the OLR patterns of figure \ref{fig:Discrete OLR} and \ref{fig:OLR_Cluster_OLR_continuous}, but the data entries refer to the OLR gradient patterns of figure \ref{fig:OLRgradcont}.} \label{tab:OLR_transitions}
\end{table}

Next we study how the patterns in figure \ref{fig:OLR_Cluster_OLR_continuous} vary in time. We construct two matrices shown in table \ref{tab:OLR_transitions}. The (i,j) element of the matrix in table \ref{tab:OLR_transitions} indicates the most frequently occurring one-day OLR anomaly pattern (depicted in figure \ref{fig:OLRgradcont}) associated with the transition from OLR spatial pattern $O_i$ on a given day to OLR spatial pattern $O_j$ on the following day. For example, row 1 of this matrix indicates the transition from OLR pattern $O_1$ to all other OLR patterns on the following day, and the diagonal indicates the maintaining of a given pattern on two consecutive days. $X$ indicate transitions that were never observed. We observe that the transition from pattern $O_1$ to $O_2$ most frequently occurs through one-day OLR anomaly pattern $OG_1$. We can summarize this symbolically as $(O_1,OG_1)\to O_2$. The number of times the transition event $(O_i,OG_k)\to O_j$ occurred, is indicated within brackets along with the corresponding matrix entry, i.e. the one-day OLR anomaly pattern. Higher numbers indicate more frequent events.

We now summarise our interpretation of the results in table \ref{tab:OLR_transitions}.
\begin{enumerate}
	\item The one-day OLR anomaly pattern $OG_3$ most often occurs when transitioning from $O_1$ into another pattern. This is reasonable, since $O_1$ represents the least convective cloud activity and any other state of OLR over India would involve an increase in convective cloud activity at all spatial locations, as represented by $OG_3$.
	\item The one-day OLR anomaly patterns $OG_6$ and $OG_8$ (representing a negative OLR gradient over most of the Indian landmass) mostly occur with OLR patterns $O_6,O_7,O_8$ (last three rows of table \ref{tab:OLR_transitions}). This suggests that the highest cloud cover patterns $(O_6,O_7,O_8)$ behave differently from the other patterns. Indeed, for the other OLR patterns, the most common one-day anomaly pattern is either $OG_4$ or $OG_5$, the patterns that depict a north-south one-day anomaly gradient.
	\item Patterns $O_6,O_7$ and $O_8$ form a family by virtue of frequent transitions (supplementary material, section 2) between themselves. Likewise patterns $O_1$,$O_2$ and $O_4$ also form a family. Note that $(O_6,O_7,O_8)$ generally coincide with more convective cloud activity whereas $(O_1,O_2,O_4)$ with lower convective cloud activity.
	\item From Table \ref{tab:Discrete OLR}, we infer that $(O_6,O_7,O_8)$ is a `high rainfall' family whereas $(O_1,O_2,O_4)$ is a `low rainfall' family. Loosely we can think of the days labeled with the former as representing an active spell, while those labeled with the latter as either a break spell, or the onset or withdrawal in June and September respectively.
	\item Pattern $O_5$ is particularly interesting since it behaves like a transition between the families of higher and lower convective cloud activity. It has almost equal likelihood between the two families. Days labeled pattern $O_5$ also occurred between active and break spells as determined by the rainfall record, using thresholds such as those used by ~\citet{Rajeevanspells}.
	\item Pattern $O_3$ is also a transition pattern, most frequently occurring after a spell of low OLR (high convective cloud activity). However, despite a lower level of rainfall (see table \ref{tab:Discrete OLR}), it is not necessarily indicative of a transition from the high convective family $(O_6,O_7,O_8)$ to the low convective family $(O_1,O_2,O_4)$ as $O_3$ has a small bias to transition into $O_6$.
	\item By and large, $OG_3,OG_4,OG_7$ represent transition into the high convective family $(O_6,O_7,O_8)$ and $OG_2,OG_5,OG_6$ represent transition into the low convective family $(O_1,O_2,O_4)$. This split is best observed in pattern $O_5$ which represents the bridge between higher and lower convective cloud activity.
	\item Evidently the one-day OLR anomaly patterns may present transitions between themselves. The most common transition for all patterns was the self-transition and the mean duration for each of them is between 1-2 days (supplementary material, section 2).
\end{enumerate}

The upshot of the above findings is that the most common means of changing the convective cloud activity across India is by establishing roughly a north-south OLR gradient. Furthermore, recall that the OLR spatial patterns depicted in figure \ref{fig:OLR_Cluster_OLR_continuous} are ordered according to increasing daily rainfall over the Indian landmass (see table \ref{tab:Discrete OLR}). Thus the above analysis may be interpreted to draw the following claims: establishing an increased convective activity in north India/monsoon zone leads, more often than not, to a period of time associated with higher rainfall overall. On the other hand, suppressing convective cloud activity in the monsoon zone/north India leads, more often than not, to a period of time associated with reduced overall rainfall. Such a relationship between OLR spatial pattern and rainfall, at this stage, is purely a statistical inference of our method. However, as pointed out in \cite{chakravarty2018}, significant deep convection is present during active phases and not present during break phases. That work was limited to the Western Ghats region but does suggest that convection has some direct relationship with precipitation. Our work is an attempt to extrapolate this inference over a wider region using the OLR satellite data. Indeed OLR patterns showing more convective cloud activity are associated with greater rainfall over the Indian landmass. In the present work we adopt this relationship to give a succinct interpretation of our findings listed above, through a probabilistic model for the one-day anomaly. The precise nature of this relationship is the object of future studies.

A similar study of dynamics can also be done with precipitation. However, the one-day anomaly patterns are less clear in this case. The results are shown in the supplementary material, section 2.

\section{Conclusion}
We have identified dominant patterns in daily OLR and rainfall in and around the Indian subcontinent by using the Markov Random Field model developed by \citet{mitra2018discrete}. We find a consistent picture emerging from three independently obtained sets of patterns: rainfall alone, OLR alone and rainfall and OLR together. Several of the patterns from one set are directly and strongly correlated to a particular pattern from another set, such as the least rainfall, highest OLR and joint lows in rainfall and convective cloud cover, which occur almost always on the same days. The monsoon days had been classified (see e.g. \citet{mitra2018discrete}) into three families: one where the entire landmass is relatively dry, another where it rains on Kerala, the Ganga basin and the north-east, and the third when it rains in Kerala and central India. From the continuous patterns emerging from the joint patterns, we find that OLR is high over most of the landmass during the first kind of day, convective cloud bands related to the second kind are oriented diagonally (from north-west to south-east direction) whereas convective cloud bands related to the third kind are aligned in the zonal direction. 

We also study the conditional relationship between OLR and rainfall, and its spatial variation across the landmass. We find all locations on the landmass to have a significant probability of  having high OLR and low rainfall even during monsoon (pre-onset, post-withdrawal and break spells), and much of southern India, except for the west coast, remains cloudy (low OLR) but dry through a significant part of the monsoon season. This is much less common in the west coast, central India and North-east, where low OLR and high rainfall often co-occur. It turns out that OLR is lowest for the rainy days in eastern India and Bay of Bengal coast and northern parts of the western coasts, suggesting that most of the rainfall occurs there from deep convective processes. But most of the rainfall in north-western India occurs under relatively high OLR.

The OLR spatial patterns that we identified are categorised into three types or families associated with low convective cloud activity, high convective cloud activity and transition pattern between high and low. The MRF-based model also determined that the OLR spatial patterns are correlated with mean rainfall over the Indian landmass: patterns associated with higher cloud convective activity were correlated with higher precipitation. Moreover, by considering patterns in the one-day anomaly we determined how the spatial patterns in convective cloud activity changed in time. Specifically, we identified a north-south OLR anomaly gradient with higher convective cloud activity in the monsoon zone, corresponded to the start of a period of higher convective cloud activity overall. Contrariwise, an anomaly with higher convective cloud activity in southern India led to a period of lower convective cloud activity overall. 

The main contribution of this paper is to utilize an MRF-based statistical model to identify a small number of static daily patterns and their temporal variation in OLR and rainfall during monsoon over India. This approach will allow us to i) visualize spatio-temporal characteristics of monsoon variables in a more comprehensive way through discrete representations, ii) potentially identify previously unknown patterns, iii) explore a framework by which the relations between multiple spatio-temporal variables can be studied, iv) develop a low-dimensional model for monsoon climate which can be used to parameterize climate models. This framework may be expanded to study other aspects of Indian monsoon such as the influences of winds, land and sea surface temperature etc, moisture transport, and also other large spatio-temporal climatic processes such as Indian Ocean Dipole, North Atlantic Dipole, El Nino-Southern Oscillation etc.

\section{Acknowledgments}
AS and AM were at ICTS-TIFR, Bangalore, India during the early stages of this work. AM acknowledges the support of the Airbus Group Corporate Foundation Chair in Mathematics of Complex Systems established in ICTS-TIFR and TIFR-CAM. We acknowledge support of the Department of Atomic Energy,
Government of India, under project no. 12-R\&D-TFR-5.10-1100. We also thank Drs. E. Suhas, N. Joseph and Prof. B. N. Goswami for providing valuable suggestions and data.

\selectlanguage{english}
\bibliographystyle{agsm}
\bibliography{converted_to_latex}

@article{rajeevan2006,
	title={High resolution daily gridded rainfall data for the Indian region: Analysis of break and active},
	author={Rajeevan, M and Bhate, J and Kale, J
	D. and Lal, B.},
	journal={Current Science},
	volume={91},
	number={3},
	pages={296--306},
	year={2006}
}

@article{mitra2018discrete,
	title={A discrete view of the Indian monsoon to identify spatial patterns of rainfall},
	author={Mitra, A and Apte, A and Govindarajan, R and Vasan, V and Vadlamani, S},
	journal={Dynamics and Statistics of the Climate System},
	volume={3},
	number={1},
	pages={1--19},
	year={2018},
	publisher={Oxford University Press US}
}

@article{mitra2018spatio,
	author = {Mitra, A and Apte, A and Govindarajan, R and Vasan, V and Vadlamani, S},
	title = "{Spatio-temporal patterns of daily Indian summer monsoon rainfall}",
	journal = {Dynamics and Statistics of the Climate System},
	volume = {3},
	number = {1},
	pages={1--16},
	year = {2019},
	month = {02},
	issn = {2059-6987},
	doi = {10.1093/climsys/dzy010},
	eprint = {http://oup.prod.sis.lan/climatesystem/article-pdf/3/1/dzy010/27765461/dzy010.pdf},
}

@article{sikka1980,
	title={On the maximum cloud zone and the ITCZ over Indian, longitudes during the southwest monsoon},
	author={Sikka, DR and Gadgil, S},
	journal={Monthly Weather Review},
	volume={108},
	number={11},
	pages={1840--1853},
	year={1980}
}

@article{schneider2014migrations,
	title={Migrations and dynamics of the intertropical convergence zone},
	author={Schneider, T and Bischoff, T and Haug, GH},
	journal={Nature},
	volume={513},
	number={7516},
	pages={45--53},
	year={2014},
	publisher={Nature Publishing Group}
}

@article{gadgil2003,
	title={The Indian monsoon and its variability},
	author={Gadgil, S.},
	journal={Annual Review of Earth and Planetary Sciences},
	volume={31},
	number={1},
	pages={429--467},
	year={2003},	
	publisher={Annual Reviews 4139 El Camino Way, PO Box 10139, Palo Alto, CA 94303-0139, USA}
}

@article{gadgil2006,
	title={The Indian monsoon, GDP and agriculture},
	author={Gadgil, Sulochana and Gadgil, Siddhartha},
	journal={Economic and Political Weekly},
	pages={4887--4895},
	year={2006},
	publisher={JSTOR}
}

@article{mishra2012,
	title={A prominent pattern of year-to-year variability in Indian Summer Monsoon Rainfall},
	author={Mishra, Vimal and Smoliak, Brian V and Lettenmaier, Dennis P and Wallace, John M},
	journal={Proceedings of the National Academy of Sciences},
	volume={109},
	number={19},
	pages={7213--7217},
	year={2012},
	publisher={National Acad Sciences}
}

@article{krishnamurthy2007,
  title={Intraseasonal and seasonally persisting patterns of Indian monsoon rainfall},
  author={Krishnamurthy, V and Shukla, J},
  journal={Journal of climate},
  volume={20},
  number={1},
  pages={3--20},
  year={2007}
}

@article{krishnamurthy2008,
	title={Seasonal persistence and propagation of intraseasonal patterns over the Indian monsoon region},
	author={Krishnamurthy, V and Shukla, J},
	journal={Climate Dynamics},
	volume={30},
	number={4},
	pages={353--369},
	year={2008},
	publisher={Springer}
}

@article{chattopadhyay2008objective,
  title={Objective identification of nonlinear convectively coupled phases of monsoon intraseasonal oscillation: Implications for prediction},
  author={Chattopadhyay, R and Sahai, AK and Goswami, BN},
  journal={Journal of the atmospheric sciences},
  volume={65},
  number={5},
  pages={1549--1569},
  year={2008}
}

@article{chattopadhyay2009,
	title={Role of stratiform rainfall in modifying the northward propagation of monsoon intraseasonal oscillation},
	author={Chattopadhyay, R and Goswami, BN and Sahai, AK and Fraedrich, K},
	journal={Journal of Geophysical Research: Atmospheres},
	volume={114},
	number={D19},
	year={2009},
	publisher={Wiley Online Library}
}

@article{Rajeevanspells,
  title={Active and break spells of the Indian summer monsoon},
  author={Rajeevan, M and Gadgil, Sulochana and Bhate, Jyoti},
  journal={Journal of earth system science},
  volume={119},
  number={3},
  pages={229--247},
  year={2010},
  publisher={Springer}
}

@article{rajeevan2013,
	title={A study of vertical cloud structure of the Indian summer monsoon using CloudSat data},
	author={Rajeevan, M and Rohini, P and Kumar, K Niranjan and Srinivasan, J and Unnikrishnan, CK},
	journal={Climate dynamics},
	volume={40},
	number={3-4},
	pages={637--650},
	year={2013},
	publisher={Springer}
}

@article{suhasMISO,
  title={An Indian monsoon intraseasonal oscillations (MISO) index for real time monitoring and forecast verification},
  author={Suhas, E and Neena, JM and Goswami, BN},
  journal={Climate dynamics},
  volume={40},
  number={11-12},
  pages={2605--2616},
  year={2013},
  publisher={Springer}
}

@article{mahakur2013high,
	title={A high-resolution outgoing longwave radiation dataset from Kalpana-1 satellite during 2004--2012},
	author={Mahakur, M and Prabhu, A and Sharma, AK and Rao, VR and Senroy, S and Singh, Randhir and Goswami, BN},
	journal={Current Science},
	pages={1124--1133},
	year={2013},
	publisher={JSTOR}
}

@article{pai2014,
  title={Development of a new high spatial resolution (0.25$\times$ 0.25) long period (1901--2010) daily gridded rainfall data set over India and its comparison with existing data sets over the region},
  author={Pai, DS and Sridhar, Latha and Rajeevan, M and Sreejith, OP and Satbhai, NS and Mukhopadhyay, B},
  journal={Mausam},
  volume={65},
  number={1},
  pages={1--18},
  year={2014}
}

@article{utsav2017,
  title={Statistical characteristics of convective clouds over the Western Ghats derived from weather radar observations},
  author={Utsav, Bhowmik and Deshpande, Sachin M and Das, Subrata K and Pandithurai, Govindan},
  journal={Journal of Geophysical Research: Atmospheres},
  volume={122},
  number={18},
  pages={10--050},
  year={2017},
  publisher={Wiley Online Library}
}

@article{chakravarty2018,
  title={Unraveling of cloud types during phases of monsoon intra-seasonal oscillations by a Ka-band Doppler weather radar},
  author={Chakravarty, Kaustav and Pokhrel, Samir and Kalshetti, Mahesh and Nair, Anish Kumar Muralidharan and Kalapureddy, Madhu Chandra R and Deshpande, Sachin M and Das, Subrata Kumar and Pandithurai, Govindan and Goswami, Bhupendra Nath},
  journal={Atmospheric Science Letters},
  volume={19},
  number={9},
  pages={e847},
  year={2018},
  publisher={Wiley Online Library}
}

@article{wang2005fundamental,
	title={Fundamental challenge in simulation and prediction of summer monsoon rainfall},
	author={Wang, Bin and Ding, Qinghua and Fu, Xiouhua and Kang, In-Sik and Jin, Kyung and Shukla, J and Doblas-Reyes, Francisco},
	journal={Geophysical Research Letters},
	volume={32},
	number={15},
	year={2005},
	publisher={Wiley Online Library}
}

@article{goswami2005enso,
	title={ENSO control on the south Asian monsoon through the length of the rainy season},
	author={Goswami, BN and Xavier, Prince K},
	journal={Geophysical Research Letters},
	volume={32},
	number={18},
	year={2005},
	publisher={Wiley Online Library}
}

@article{pottapinjara2016relation,
	title={Relation between the upper ocean heat content in the equatorial Atlantic during boreal spring and the Indian monsoon rainfall during June--September},
	author={Pottapinjara, Vijay and Girishkumar, MS and Sivareddy, S and Ravichandran, M and Murtugudde, R},
	journal={International Journal of Climatology},
	volume={36},
	number={6},
	pages={2469--2480},
	year={2016},
	publisher={Wiley Online Library}
}

\end{document}


\maketitle
\selectlanguage{english}

\section{Composite Analysis of the Spatial Patterns}\label{sec:Composite}
\begin{figure}[h!]
	\centering
	\includegraphics[width=0.8\textwidth]{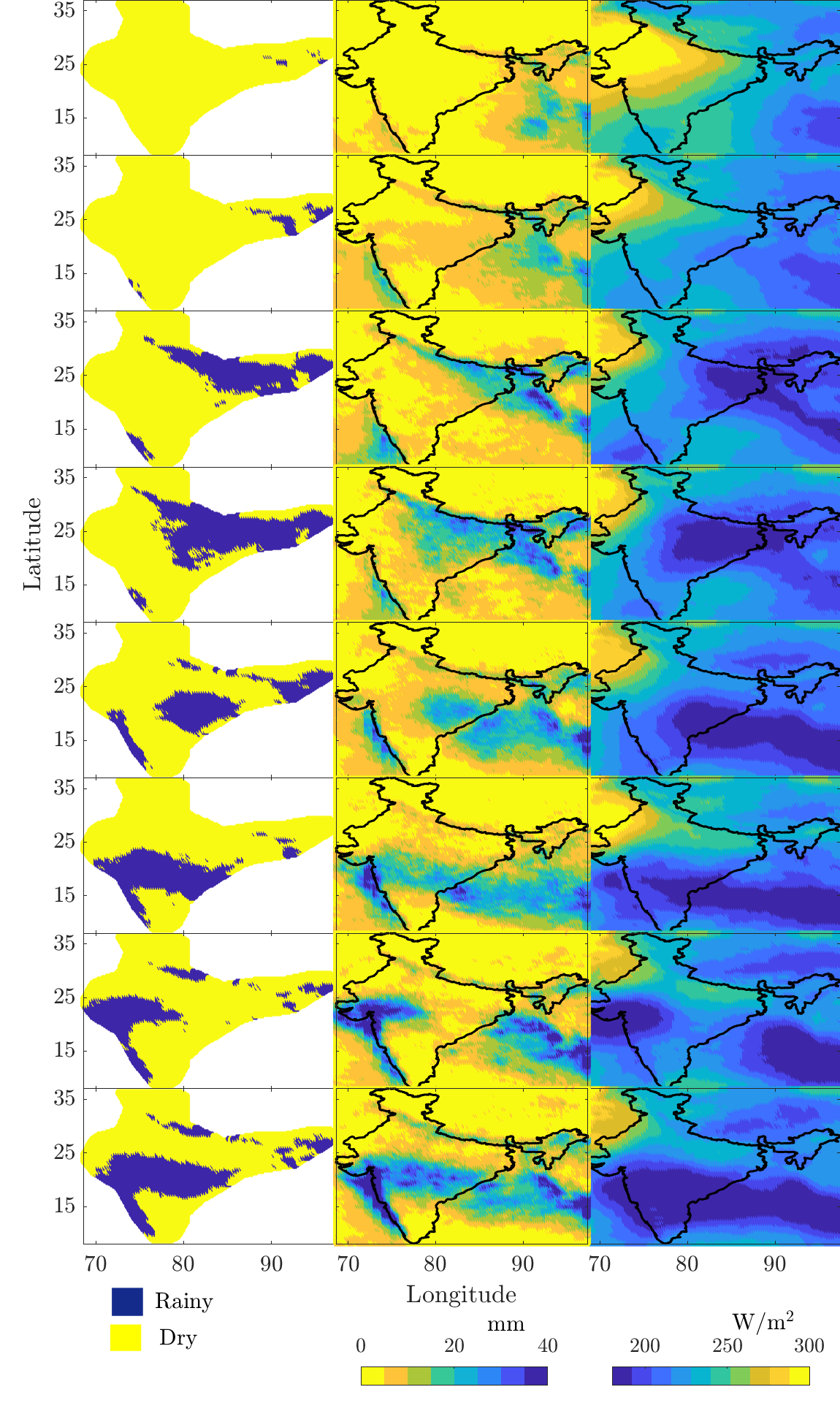}
	\caption {Composites associated with the rainfall patterns. Left, middle and right column represent the discrete rainfall pattern, continuous rainfall and continuous OLR pattern.\label{fig:PPT_Composite}}
\end{figure}

\begin{figure}[h!]
	\centering
	\includegraphics[width=0.8\textwidth]{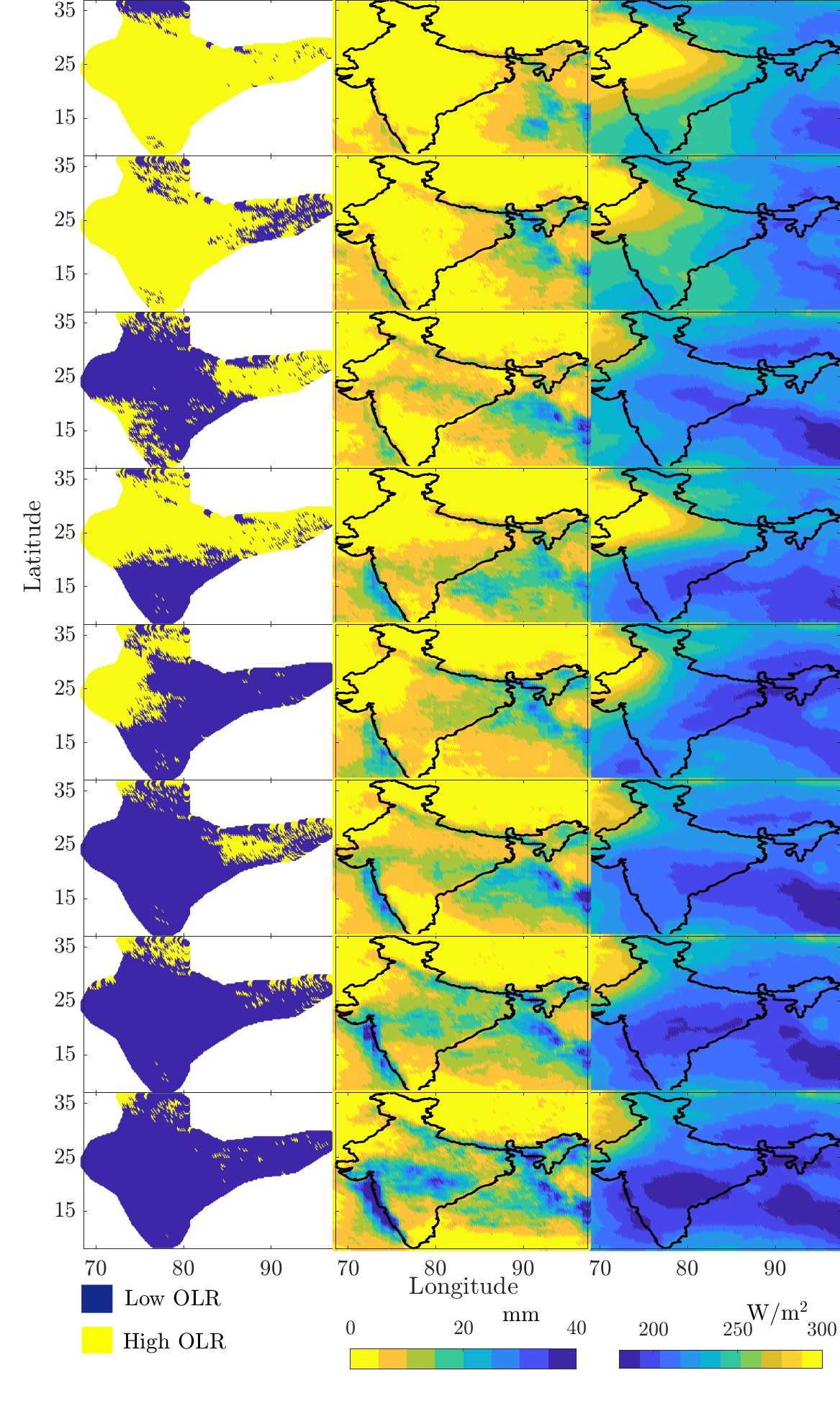}
	\caption {Composites associated with the OLR patterns. Left, middle and right column represent the discrete OLR pattern, continuous rainfall and continuous OLR pattern.\label{fig:OLR_Composite}}
\end{figure}

\begin{figure}[h!]
	\centering
	\includegraphics[width=0.8\textwidth]{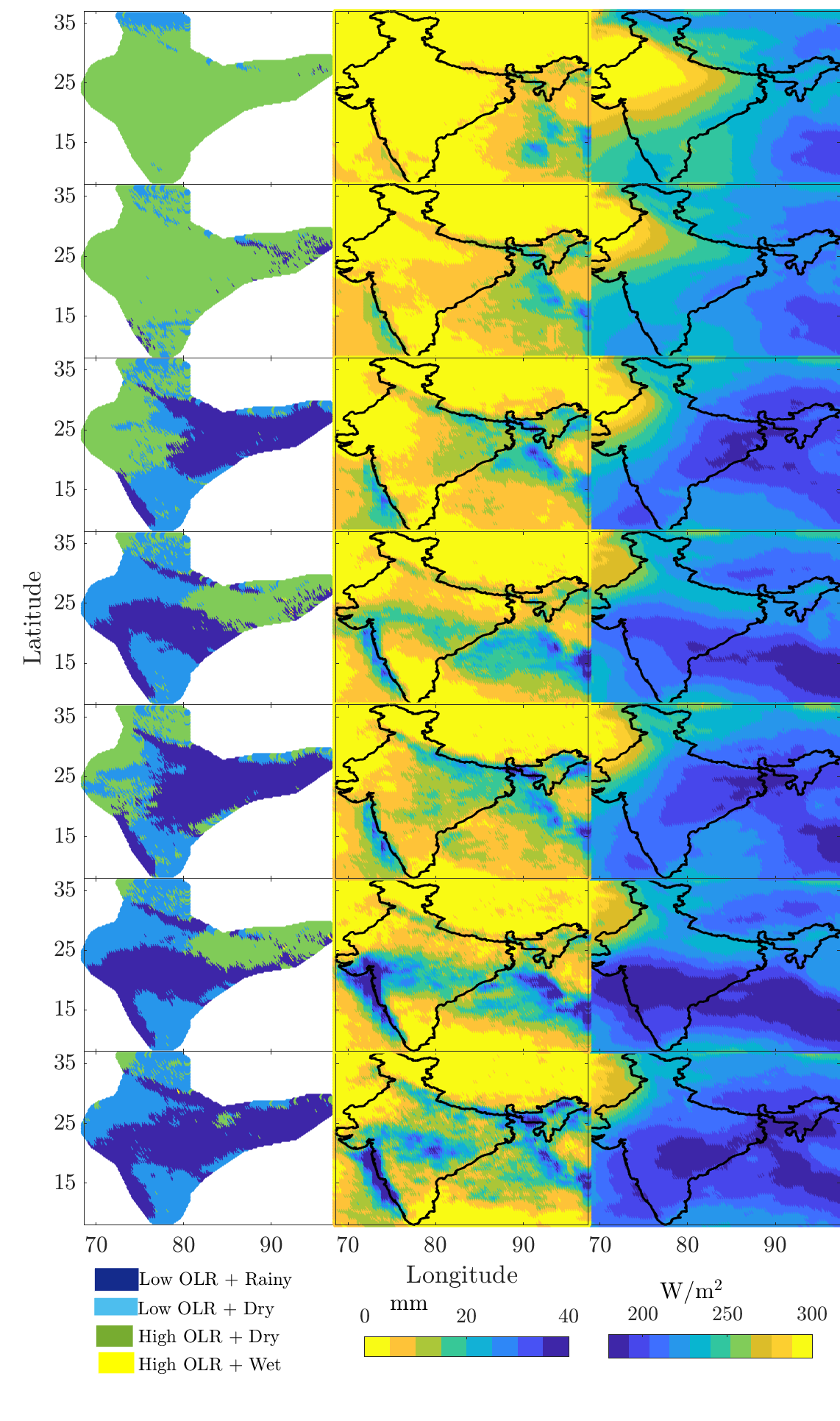}
	\caption {Composites associated with the joint patterns. Left, middle and right column represent the discrete joint pattern, continuous rainfall and continuous OLR pattern.\label{fig:Joint_Composite}}
\end{figure}

In section 3 of the main paper, we identify several spatial patterns of rainfall and OLR. For a more comprehensive understanding between OLR and rainfall, we visualize these patterns through composite figures. We compile the mean rainfall and OLR over the days associated with each such pattern. The composites for the rainfall (R), OLR(O) and joint (J) patterns shown in figures 1, 3 and 6 of the main paper are shown here in figures \ref{fig:PPT_Composite}, \ref{fig:OLR_Composite} and \ref{fig:Joint_Composite}. 


\section{Temporal variation in precipitation spatial patterns}\label{sec:AppendixTemporalGrad}
\begin{figure}[h!]
	\centering
	\includegraphics[width=5.5in, height=3in]{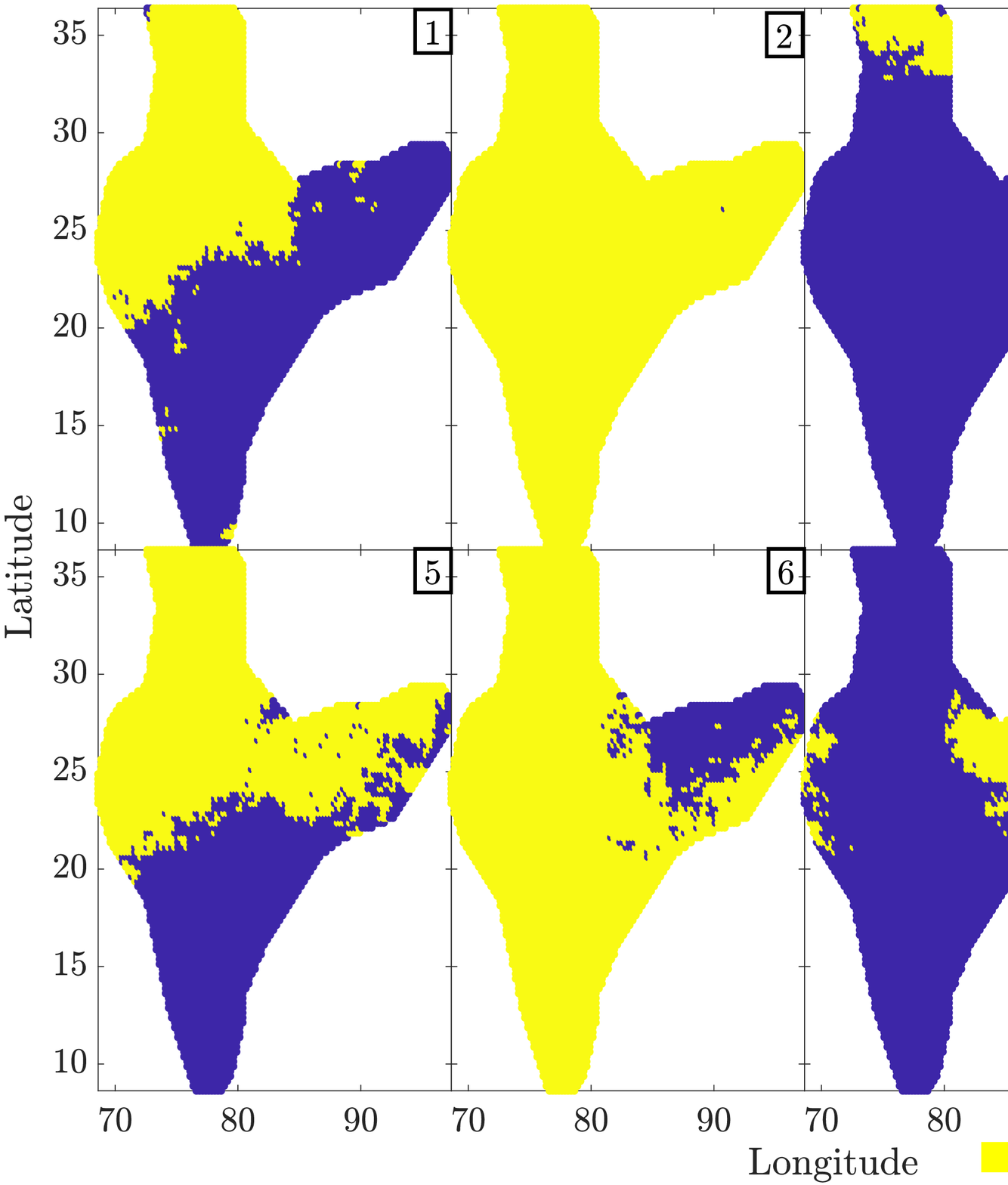}
	\caption {Discrete spatial patterns corresponding to the temporal gradient data of OLR, during the period 2004-10. Yellow indicates an increase in OLR (less cloud on the following day), blue indicates a decrease in OLR (more cloud on the following day).\label{fig:OLRgrad}}
\end{figure}

In section 4  of the main paper we discussed the day-to-day variation of OLR  using the continuous patterns of temporal gradients or one-day anomalies. The discrete patterns corresponding to those continuous patterns in figures 13 of the main paper are shown in figure \ref{fig:OLRgrad}.

\begin{figure}[h!]
	\centering
	\subfloat{\includegraphics[width=0.5\textwidth]{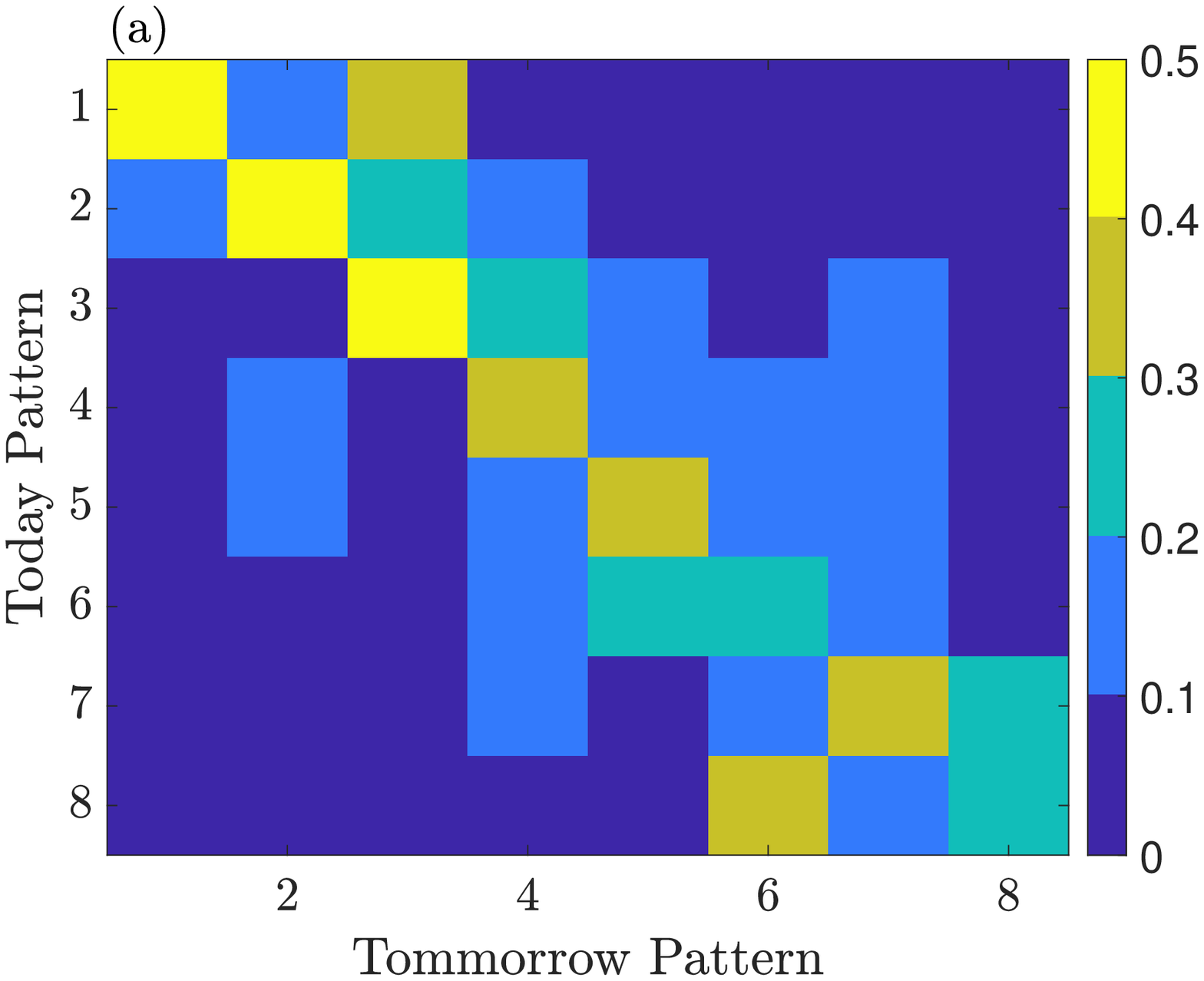}}\hfill
	\subfloat{\includegraphics[width=0.5\textwidth]{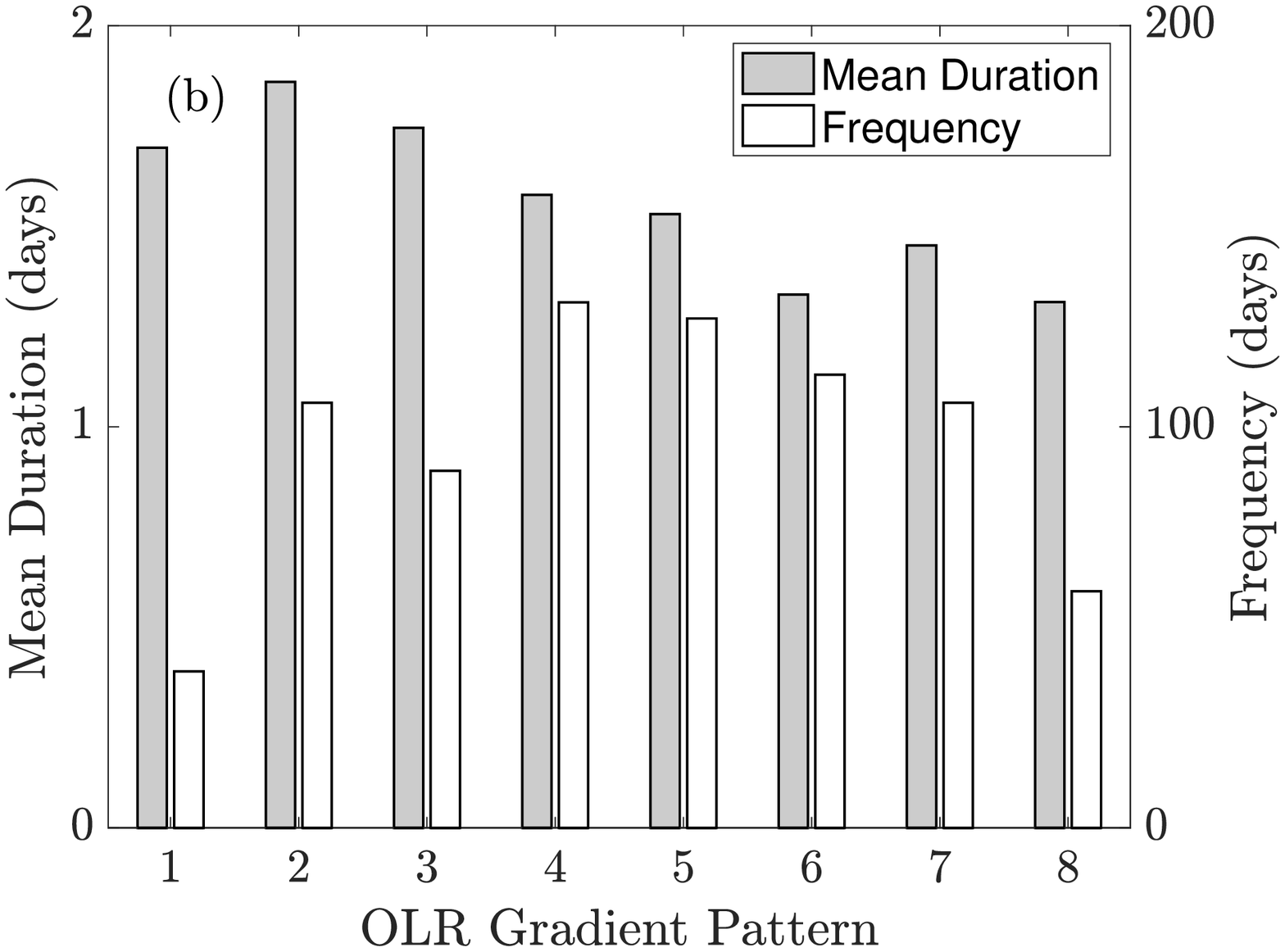}}
	\caption {Statistics related to the one -day OLR anomaly patterns: a) Transition matrix and b) Mean duration and Residence time. \label{fig:OLRgradstats}}
\end{figure}

The transition matrices for one-day OLR anomaly patterns along with the mean duration (residence time) and frequency of each patterns are shown in figure~\ref{fig:OLRgradstats}.  These figures are the basis of the interpretation in Section 4 of the main paper. 

\begin{figure}[h!]
	\centering
	\includegraphics[width=5.5in, height=3in]{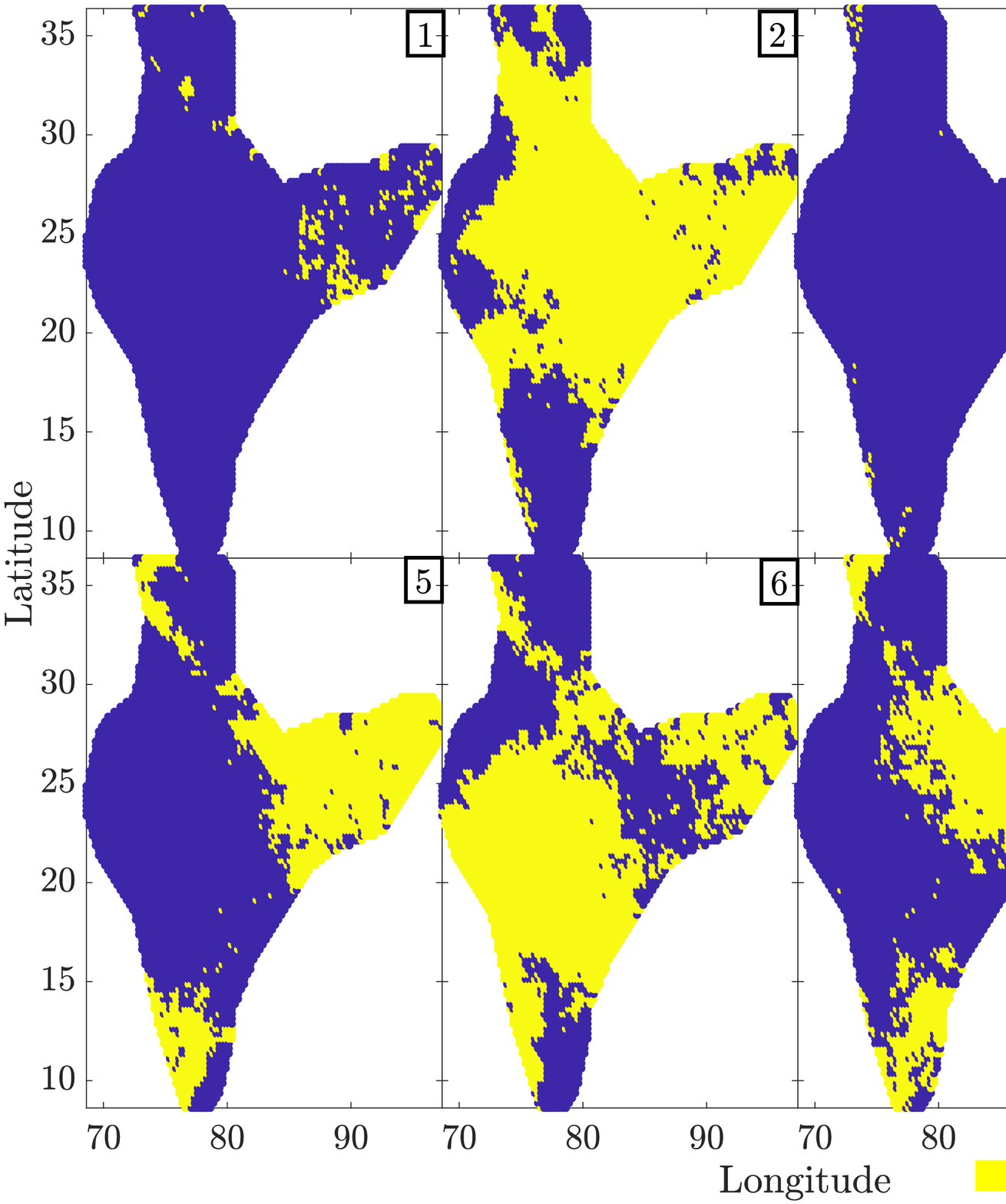}
	\caption {Discrete spatial patterns corresponding to the temporal gradient data of precipitation, during the period 2004-10. Yellow indicates a decrease in rainfall, blue indicates an increase in rainfall. \label{fig:PPTgrad}}
\end{figure}

\begin{figure}[h!]
	\centering
	\includegraphics[width=5.5in, height=3in]{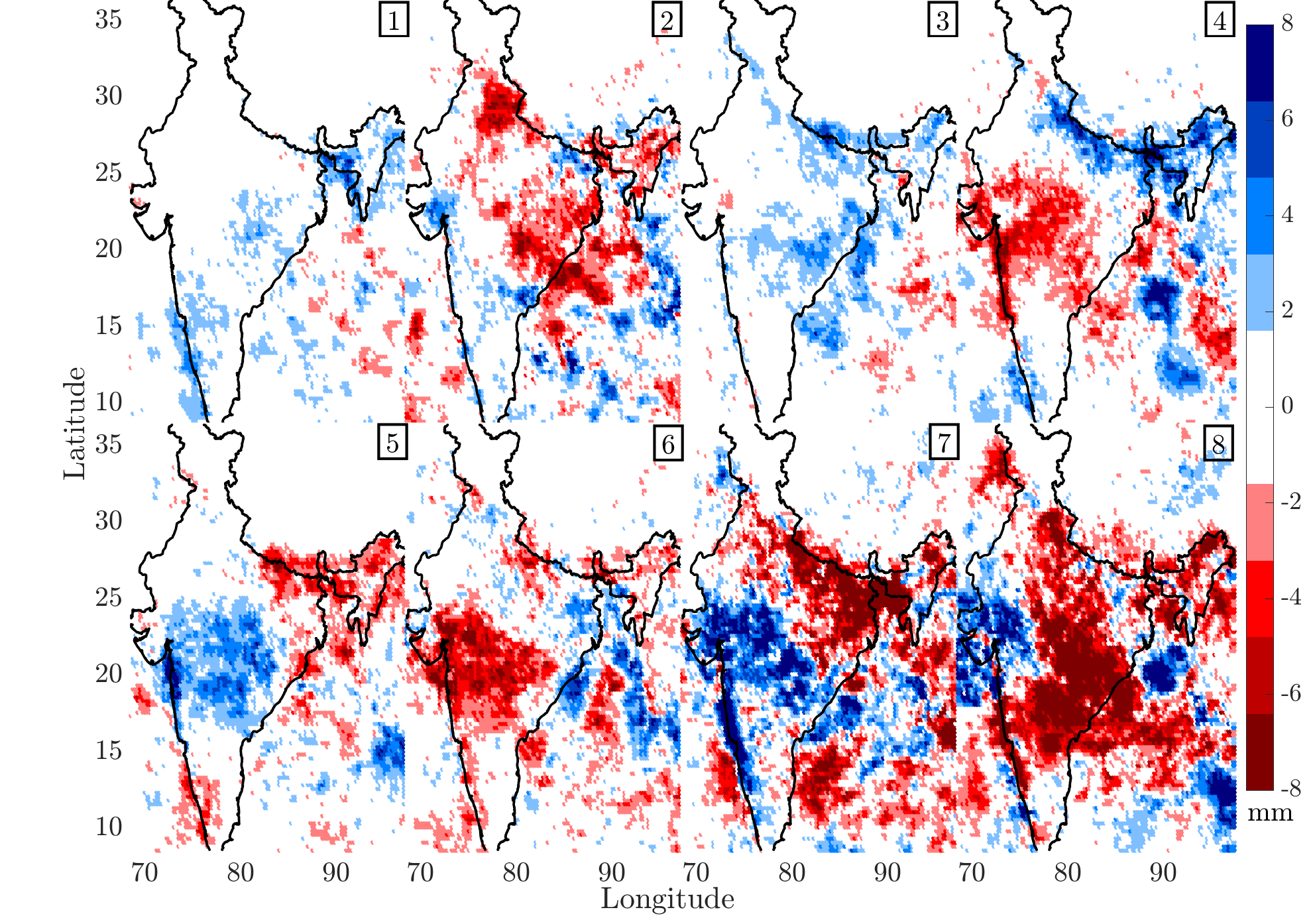}
	\caption {Continuous spatial patterns corresponding to the temporal gradient data of precipitation, during the period 2004-10. Red indicates a decrease in rainfall, blue indicates an increase in rainfall. \label{fig:PPTgradcont}}
\end{figure}

We have also repeated the analysis of the one-day rainfall anomaly and show the continuous one-day rainfall anomaly patterns in Figure \ref{fig:PPTgradcont} and the discrete counterparts in figure \ref{fig:PPTgrad} with relevant statistics in table~\ref{tab:Discrete_PPTgrad}. 

\begin{table}[h]                           
	\centering
	\begin{tabular}{|c|c|c|c|c|c|c|c|c|}
		\hline
		PPT gradient pattern number    &   RG1 & RG2& RG3 &RG4& RG5& RG6& RG7 & RG8               \\ \hline
		Number of days &  145 & 39& 198 &107& 158& 105& 49 & 46           \\ \hline    
		Mean landmass rainfall  (mm/day) &  4.40 & 4.8& 6.5 &7.1& 7.9& 8.8& 9.6 &9.9 \\ \hline    
	\end{tabular}
	\caption{Number of attributed days and daily mean rainfall for the discrete PPT temporal gradient patterns of figure \ref{fig:PPTgrad}. These rainfall gradient patterns together account for 847 of the 854 days.} \label{tab:Discrete_PPTgrad}                            
\end{table}

\section{Propagation based on CoM analysis}\label{sec:Propagation}

\begin{figure}[h!]
	\centering
	\subfloat{\includegraphics[width=0.33\textwidth]{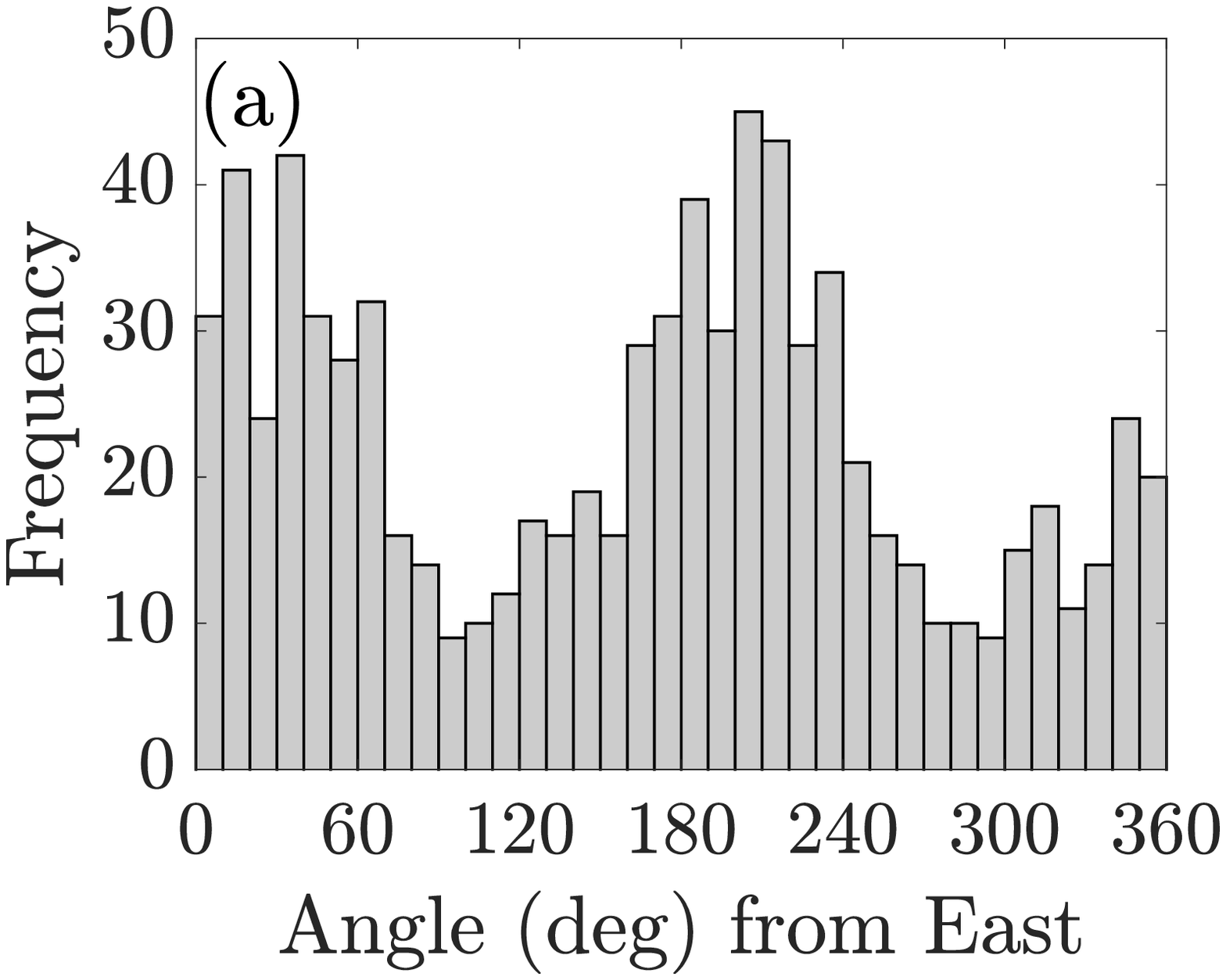}\label{fig:Histogram_angles_landmass_rain_raw}}\hfill
	\subfloat{\includegraphics[width=0.33\textwidth]{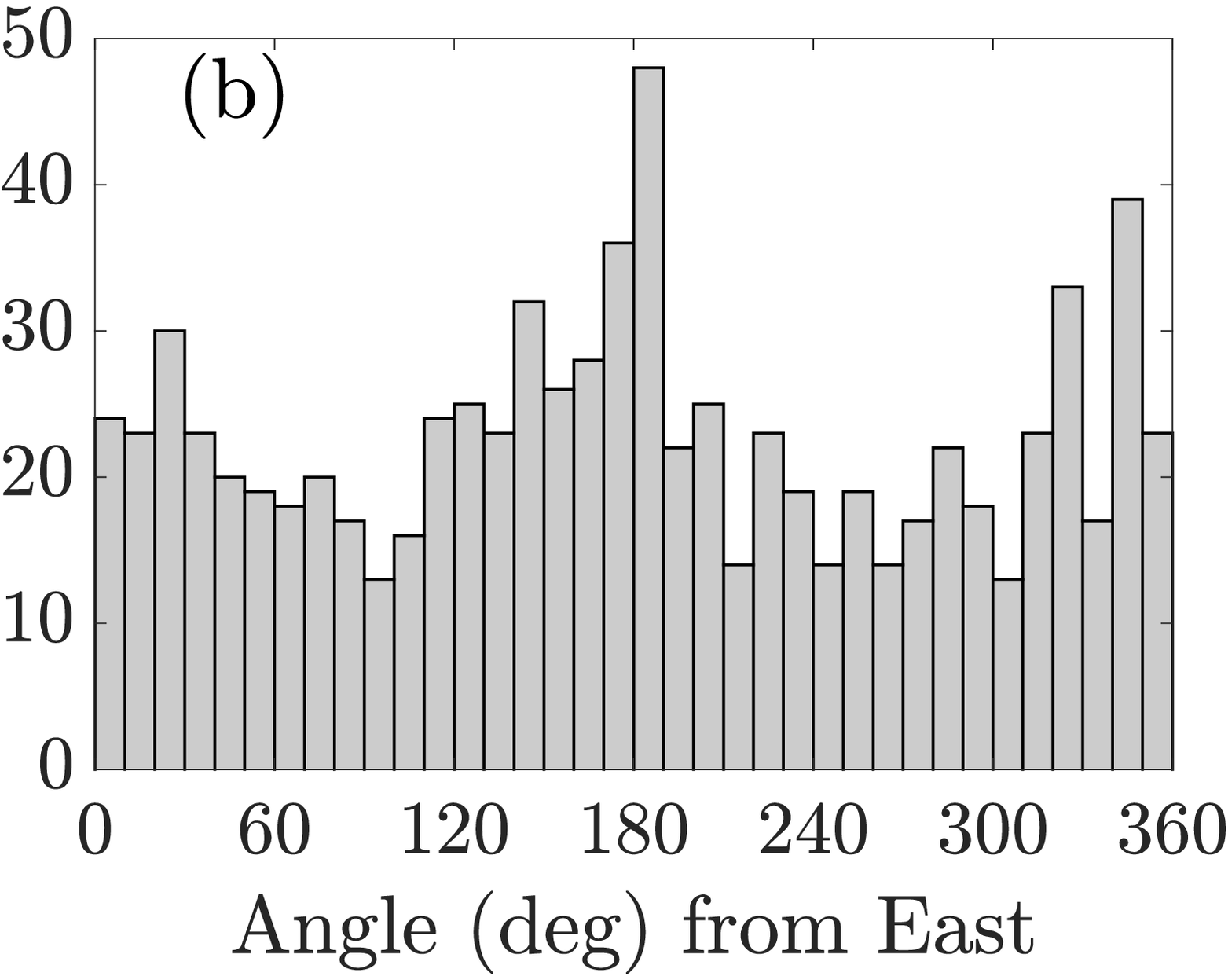}\label{fig:Histogram_angles_extended_rain_raw}}\hfill
	\subfloat{\includegraphics[width=0.33\textwidth]{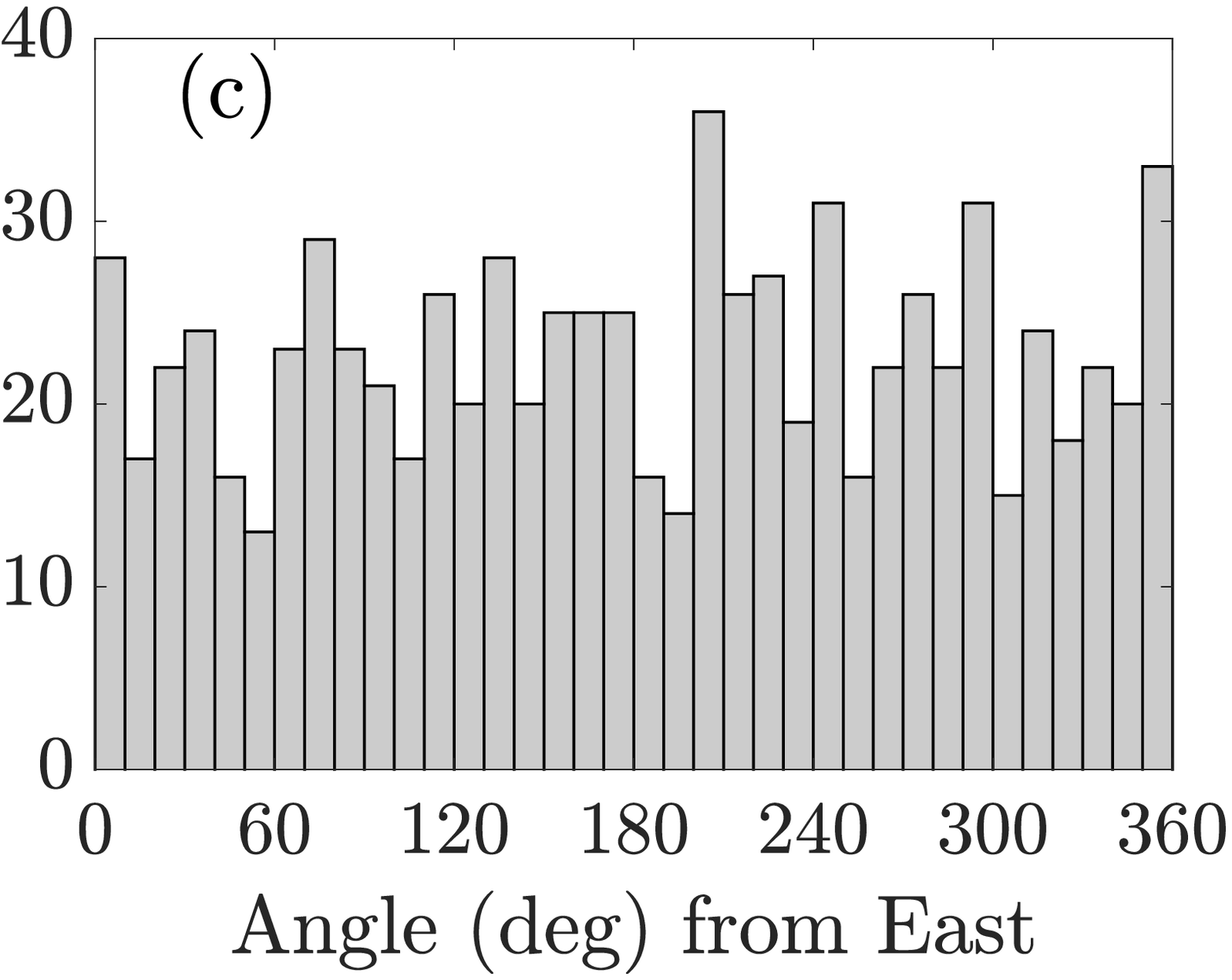}\label{fig:Histogram_angles_extended_OLR_raw}}
	\caption {The direction of the velocity vector of Weighted Centre of Mass (Def. 1) for (a) Landmass Rainfall, (b) Extended area Rainfall and (c) Extended area OLR.  The velocity vector of the landmass rain c.o.m. in figure \ref{fig:Histogram_angles_landmass_rain_raw} shows a bimodal distribution, with its angle clustered between 0 and 60 degrees (defining due east as 0 degrees) and between 160 and 240 degrees. This indicates that maximum propagation of rainfall within the landmass takes place along the north-east or south-west directions. However, the velocity vector of the extended rain c.o.m. is seen in figure \ref{fig:Histogram_angles_extended_rain_raw} to be more uniformly distributed across all angles with a slight but distinct peak between 170 and 250 degrees (i.e. south-west). Similarly, nothing specific can be said about cloud propagation as the histogram of c.o.m. of 1/OLR in figure \ref{fig:Histogram_angles_extended_OLR_raw} is also close to uniform. \label{fig:Hists_raw}}
\end{figure}

\begin{figure}[h!]
	\centering
	\subfloat{\includegraphics[width=0.33\textwidth]{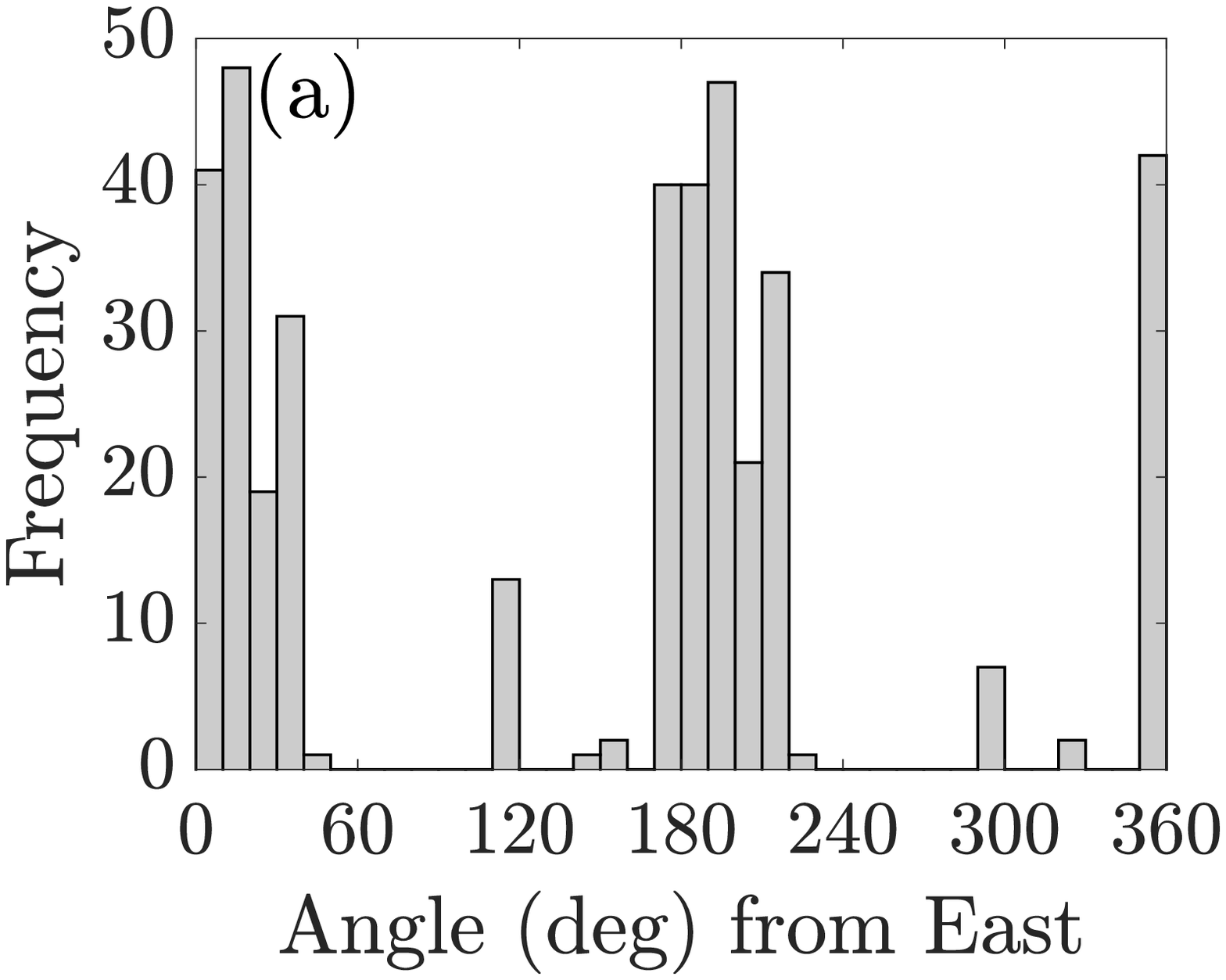}\label{fig:Histogram_angles_landmass_rain_joint_clusters}}\hfill
	\subfloat{\includegraphics[width=0.33\textwidth]{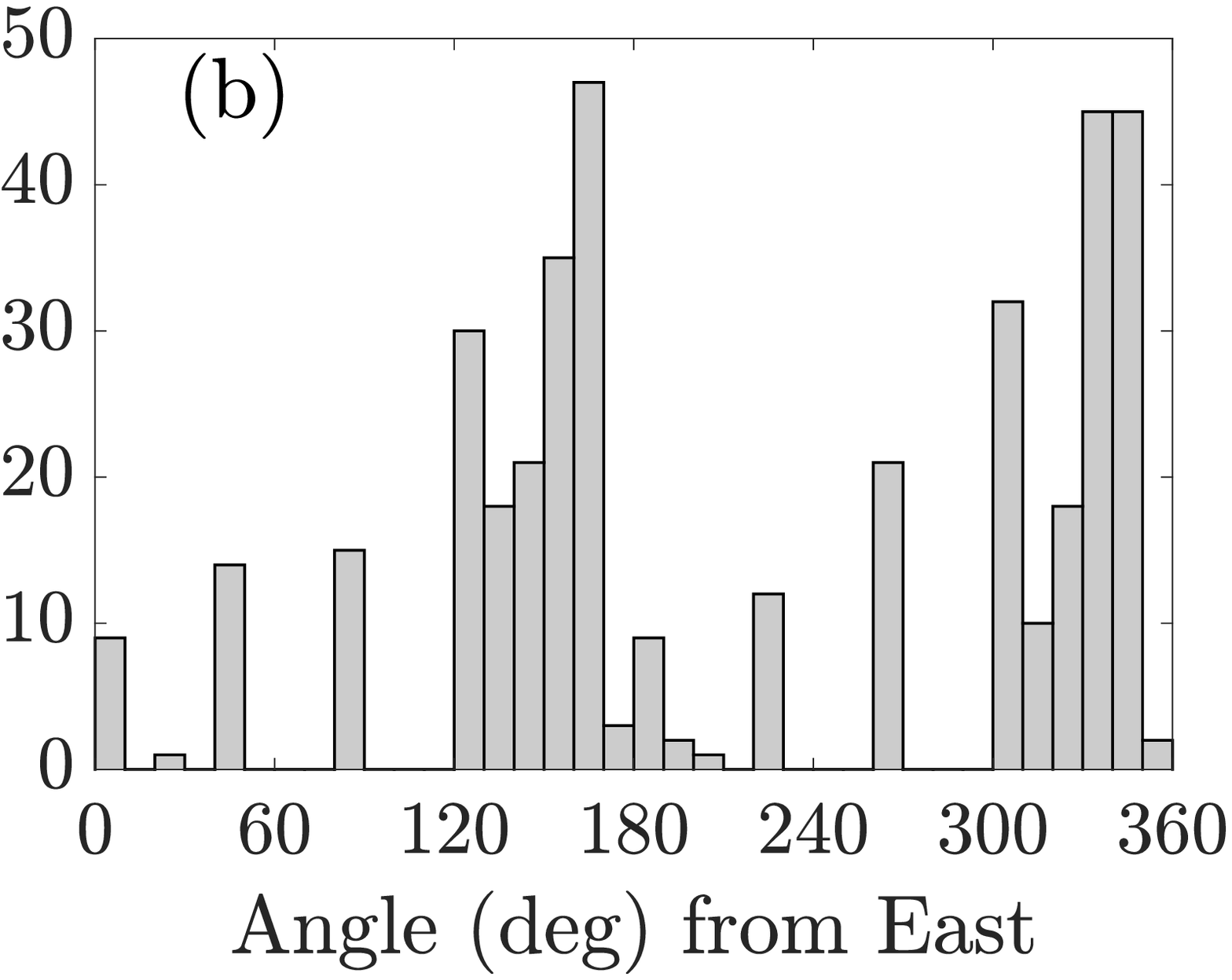}\label{fig:Histogram_angles_extended_rain_joint_clusters}}\hfill
	\subfloat{\includegraphics[width=0.33\textwidth]{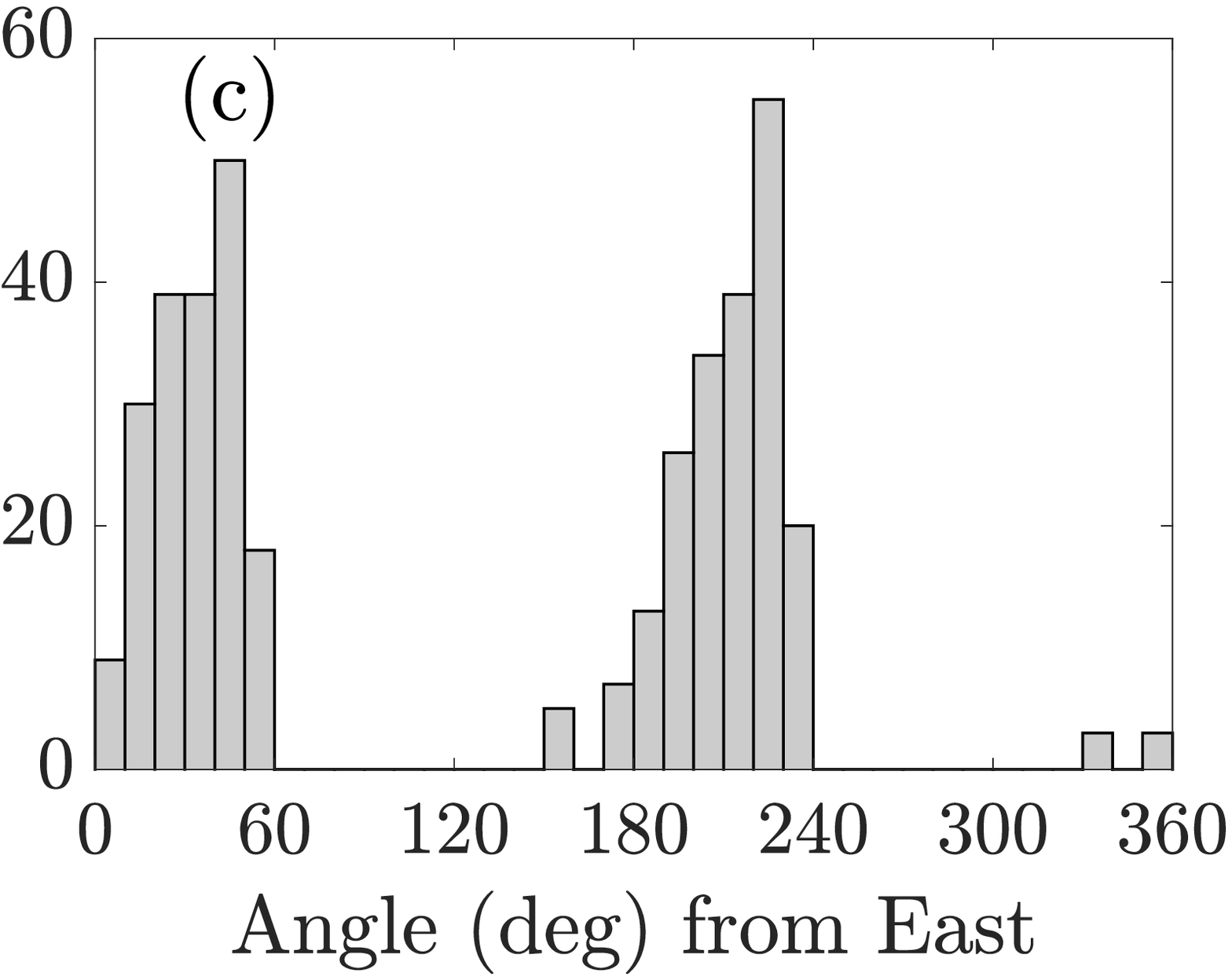}\label{fig:Histogram_angles_extended_OLR_joint_clusters}}
	\caption {The direction of the velocity vector of Centre of Mass (Def. 1) of continuous joint patterns for (a) Landmass Rainfall, (b) Extended area Rainfall and (c) Extended area OLR. The velocity vector of the c.o.m. of rain over the Indian landmass is peaked around 10 and 200 degrees, with almost no presence at any other angle, in figure \ref{fig:Histogram_angles_landmass_rain_joint_clusters}. On the other hand, peaks in the velocity of the extended landmass rainfall c.o.m. are seen at angles around 340 and 160 degrees in figure \ref{fig:Histogram_angles_extended_rain_joint_clusters}. These are quite close to the angles on the Indian landmass, both showing practically a zonal propagation of the centre of mass of rain. The movement of landmass rainfall is along the east-north-east direction whereas it is along the east-south-east direction for extended area rainfall. On the other hand, the velocity of the c.o.m. of 1/OLR peaks around 40 and 220 degrees in figure \ref{fig:Histogram_angles_extended_OLR_joint_clusters}. The two peaks in the angle distribution of velocity for each type of c.o.m. indicate two distinct phases of the rainfall or cloud movements.\label{fig:Hists_joint}}
\end{figure}

\begin{figure}[h!]
	\centering
	\includegraphics[width=1\textwidth]{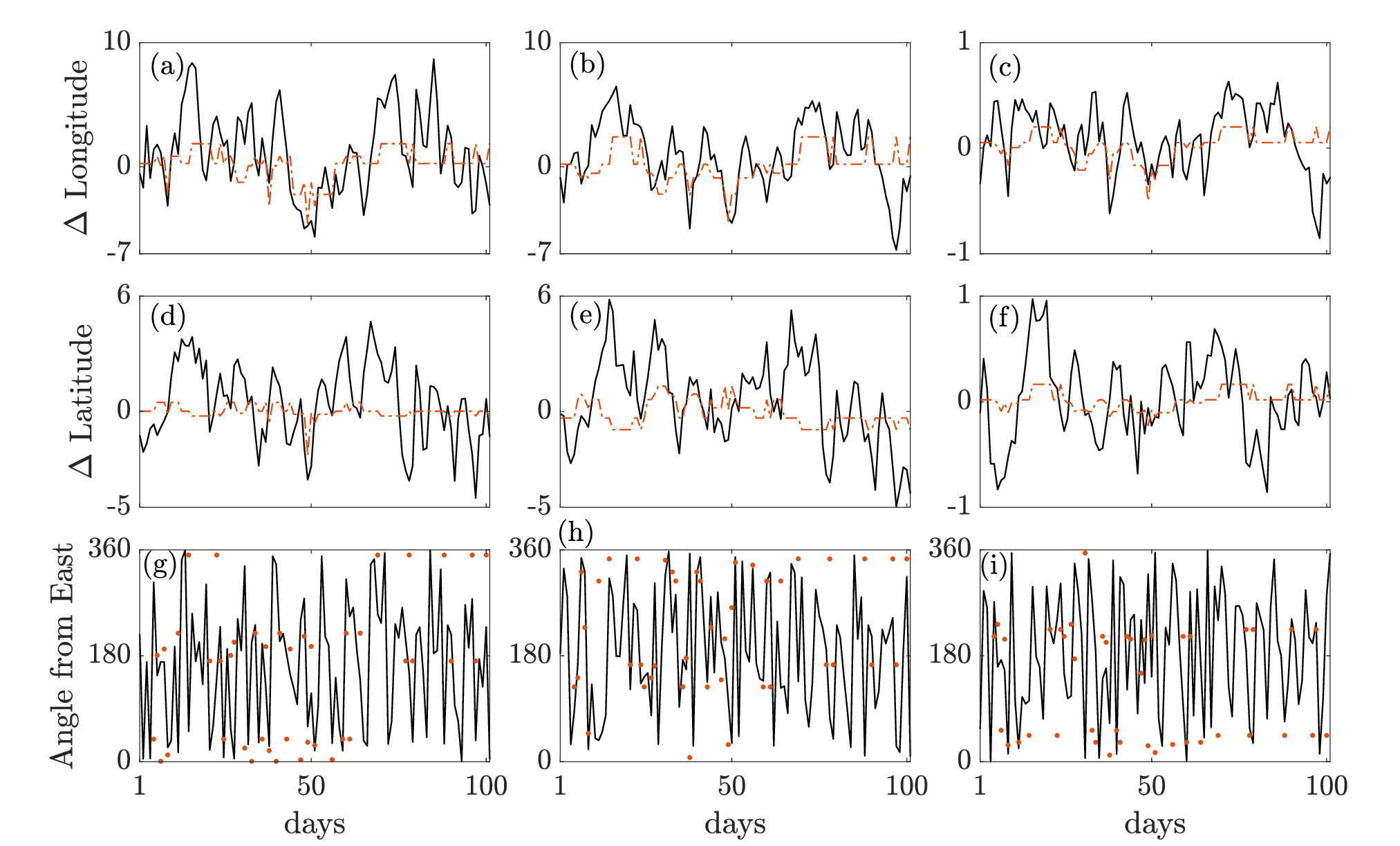}
	\caption {Time series (daily) of the deviation of longitude (top row) and latitude (bottom row) from their average and the velocity angle from the east (bottom row) for the centre of mass (c.o.m.) (Def. 1) weighted with landmass rainfall (left column), extended area rainfall (middle column) and extended area 1/OLR (right column) from 6 June 2004 to 30 September 2004. The solid black line is from the raw data and the orange broken line or markers are from the joint pattern. When a given pattern is persistent over more than one day (zero velocity of c.o.m.) the velocity angle from the pattern cannot be defined and hence markers have been used in the bottom row. Instead of using the raw data if we use the continuous pattern obtained from joint clustering associated with each day we obtain a clearer picture, as shown in figure \ref{fig:Hists_joint}. The movement of clouds is thus either towards the north-east or towards the south-west. We find the 10-degree movement of landmass rainfall c.o.m., 340-degree movement of the extended area c.o.m. and the 40-degree movement of the 1/OLR c.o.m. to coincide on most occasions. The 200, 160 and 220 degrees movement of these c.o.m.s also coincides. \label{fig:Time_Series_latitudes_angles}}
\end{figure}

To deduce the movement of the cloud bands, and their relationship to the evolution of the rainfall distribution, we calculate the Centre of Mass (c.o.m.) of the rainfall and the inverse OLR (1/OLR) for each day (where rainfall and 1/OLR values at a given location form the weights). We consider two definitions of the location ($Y_{com}(t)=[lat_{com}(t), lon_{com}(t)]$) of centre of mass (CoM) on each day:  \vspace{-0.2in}
\begin{enumerate}
	\item mean of the geospatial coordinates (latitude and longitude) within the area, weighted by volume of rainfall or 1/OLR,
	\begin{equation}\label{eq:Def1}
	lat_{com}(t)=\frac{\sum_{s}X(s,t)lat(s)}{\sum_{s}X(s,t)}, \quad lon_{com}(t)=\frac{\sum_{s}X(s,t)lon(s)}{\sum_{s}X(s,t)},
	\end{equation}
	\item mean of the geospatial coordinates (latitude and longitude) within the area, weighted by the binary state of rainfall or 1/OLR as calculated by the model,
	\begin{equation}\label{eq:Def2}
	lat_{com}(t)=\frac{\sum_{s}Z(s,t)lat(s)}{\sum_{s}Z(s,t)}, \quad lon_{com}(t)=\frac{\sum_{s}Z(s,t)lon(s)}{\sum_{s}Z(s,t)}.
	\end{equation}
\end{enumerate}
The velocity of CoM, with components in the meridional and zonal direction, on each day is defined as $V_{com}(t) = Y_{com}(t)-Y_{com}(t-1)$. The statistics related to the angle of velocity vector are shown in figures \ref{fig:Hists_raw}, \ref{fig:Hists_joint} and \ref{fig:Time_Series_latitudes_angles}. Movement of both convective clouds and rainfall is primarily restricted to a dominant direction, and the two differ by about 30-40 degrees. 

